\documentclass{jpp}

\usepackage{amsmath,amssymb,bm}
\allowdisplaybreaks[3]
\usepackage{siunitx}
\usepackage{graphicx}\graphicspath{{}{Graph/}{Graph/seminar/}
    {./Graph/}{./Graph/A0/}{./Graph/A1/}{./Graph/A2/}{./Graph/A3/}{./Graph/An/}
}

    \usepackage[svgnames]{xcolor}

\usepackage{iftex}
\iftutex
    \usepackage{fontspec}

    \usepackage[math-style=ISO,bold-style=ISO]{unicode-math}
\else

    \usepackage[T1]{fontenc} 
    \usepackage[utf8]{inputenc} 
\fi

    \iftutex
        
    \else
        
    \fi

    \DeclareMathOperator{\const}{const}

    \DeclareMathOperator{\Root}{Root}

    \newcommand{\tparder}[2]{{\partial #1}/{\partial #2}}
    \newcommand{\parder}[2]{\frac{\partial #1}{\partial #2}}

    \newcommand{\dif}[2][]{\mathop{}\!\mathrm{d}
        \if
            \relax\detokenize{#1}\relax
        \else
            ^{\mkern-1.mu#1}\mkern-2.5mu 
    \fi
    #2\,}
    \newcommand{\der}[2]{\frac{\dif{#1}}{\dif{#2}}}
    \newcommand{\tder}[2]{{\dif{#1}}/{\dif{#2}}}

    \newcommand{\mean}[1]{\left\langle #1\right\rangle}

\usepackage[
    colorlinks
        ,linkcolor=violet
        ,filecolor=purple
        ,citecolor=teal
        ,urlcolor=magenta
    ,unicode
    ,pdfpagemode=UseOutlines
    ,pdfdisplaydoctitle=true
    ,pdfpagetransition=Wipe
    ,pdfusetitle
]{hyperref}
\hypersetup{
        pdfkeywords={plasma, MHD stability, ballooning modes, LoDestro equation, mirror trap, Gas-Dynamic Trap, GDT, Compact Axisymmetric Toroid, CAT, Wisconsin HTS Axisymmetric Mirror, WHAM}
}
\usepackage{bookmark}

\begin{document}

\def\bot{\mathrel\perp}

\shorttitle{On the stability of rigid ballooning mode}
\shortauthor{I. A. Kotelnikov}

\title{
    On the stability 
    \\of the $m=1$ rigid ballooning mode 
    \\in a mirror trap with high-beta sloshing ions
}

\author{Igor A. Kotelnikov\aff{1}
  \corresp{\email{igor.kotelnikov@gmail.com}}}

\affiliation{
    \aff{1}Novosibirsk State University, Novosibirsk, Russia
    \aff{2}Budker Institute of Nuclear Physics, Novosibirsk, Russia}

\begin{abstract}
Stability of the “rigid” $m = 1$ ballooning mode (RBM) in a mirror axisymmetric trap is studied for the case of oblique neutral beam injection (NBI), which creates an anisotropic  population of fast sloshing ions. Since small-scale modes with azimuthal numbers $m>1$ in long thin (paraxial) mirror traps are easily stabilized by finite Larmor radius (FLR) effects, suppression of the rigid ballooning and flute modes would  mean stabilization of all MHD modes, with the exception of mirror and fire hose disturbances, which are intensively studied in geophysics, but have not yet been identified in mirror traps.

Large-scale ballooning mode can, in principle, be suppressed either by the lateral perfectly conducting wall, or by the end MHD anchors (such as cusp or by biased limiters), or by their combination. The effect of the wall shape, vacuum gap width between the plasma column and the lateral wall, angle of oblique NBI, radial profile of the plasma pressure, and axial profile of the vacuum magnetic field are studied.

It is confirmed that the lateral conducting wall still creates the upper stability zone $\beta>\beta_{\text{cr2}}$. However, in many cases the upper zone is clamped from above by the mirror or fire hose instabilities. When the lateral wall is combined with end MHD anchors, a lower stability zone $\beta<\beta_{\text{cr}1}$ appears. These two zones can overlap in case of a sufficiently smooth radial pressure profile, and/or a sufficiently low mirror ratio, and/or a sufficiently narrow vacuum gap between the plasma column and the lateral wall. However, even in this case, the range of permissible values of beta is limited from the above  either by the threshold of mirror instability $\beta_{\text{mm}}<1$ or by the threshold of fire hose instability $\beta_{\text{fh}}<1$, in contrast to the case of transverse NBI, where plasma pressure is limited by the condition $\beta <1$ that guaranties existence of plasma confinement by external magnetic field.  

\end{abstract}

\keywords{MHD stability, ballooning modes, LoDestro equation, mirror trap}

\maketitle

\section{Introduction}\label{s1}



In a series of articles, stability of a so-called rigid ballooning mode an azimuthal number $m=1$ was studied for several model configurations of an axisymmetric mirror trap (also called open or linear traps) was studied using a numerical solution of the LoDestro equation \citep{LoDestro1986PF_29_2329}. The conditions for stabilization of these modes were found assuming both a lateral perfectly conducting cylindrical wall surrounding the plasma column and traditional end MHD stabilizers attached to the central cell of such a trap. A few model radial and axial plasma pressure profiles and model axial vacuum magnetic field profiles we examined. 

In quite realistic experimental conditions, oscillations of a ballooning perturbation with an azimuthal number $m\geqslant 2$ are easily stabilized by finite-Larmor-radius effects (FLR effects), which, however, cannot stabilize the $m=1$ mode, although make it ``rigid'' in a certain sense. Thus, stabilization of the rigid ballooning mode $m=1$ is, in practice, equivalent to the stabilization of all types of ballooning and flute disturbances.

In the first article \citep{Kotelnikov+2022NF_62_096025}, which was followed by the two more papers \citep{Kotelnikov+2023NF_63_066027,ZengKotelnikov2024PPCF_66_075020}, a plasma model with isotropic pressure was implemented. It was assumed that the plasma column is surrounded by a cylindrical perfectly conducting chamber with a variable radius, which is proportional to the plasma radius with a constant coefficient. This shape was later called ``Proportional'' because it reproduced the shape of the plasma column on a larger scale. 

The numerical code was then upgraded to implement a transverse neutral beam injection model (transverse NBI), which simulates the anisotropic fast ion pressure that occurs when a beam of fast neutral atoms is injected into a relatively cold target plasma at right angles to the axis trap at a magnetic field minimum \citep{Kotelnikov+2023NF_63_066027}. 

The next update to the numeric code added the ability to select a custom conductive chamber shape. This possibility was demonstrated assuming an example of a conducting chamber in the shape of a straight cylinder. Calculations published in \citet{ZengKotelnikov2024PPCF_66_075020} made it possible to compare the stabilizing effect of a proportional conducting chamber with the effect of such a ``straightened'' chamber for a model of a transverse NBI.

The present paper reports the results of study for the case on oblique NBI. Oblique NBI yields a population of fast ions, which are called sloshing ions. They are characterized by a pressure distribution with a peak located somewhere between the median plane of the mirror trap and magnetic mirror plugs. It is significant that, in addition to the ballooning instability, mirror and fire hose-type instabilities can also be raised under oblique injection. It will be shown below that in the high mirror ratio trap, it is these instabilities that determine the ultimate beta limit.

The present article completes the series of three above cited publications that used model functions to approximate  radial and axial profiles of the plasma pressure and vacuum magnetic fields. They were convenient for identifying general patterns in how the ballooning instability threshold and margin depend on the mirror ratio, width of the magnetic mirror plugs, shape and steepness of the radial and axial plasma pressure profile, size and shape of the vacuum gap between the lateral conducting wall and the lateral surface of the plasma column, as well as upon the stability margin created due to the end MHD anchors such as magnetic cusps.

Each stage began with a code modernization, and each modernization was accompanied by a complication of the numerical code and an increase in the duration of calculations. If for an isotropic plasma the recalculation of the data necessary to reproduce all the graphs published so far took no more than an hour on a desktop computer with 4 to 8 processors of the Intel i7-Core level, then in the case of a transverse NBI and a straightened camera, this would take few weeks. The oblique NBI calculations presented in this paper initially took few months of intensive computations intermixed with code development and testing. And only the transfer of calculations to a multi-core cluster made it possible to return the duration of calculations to a reasonable range of few days.

The reason for the slowdown in numerical code is trivial: with each update, the number of integrations that could be performed analytically inevitably decreased.

In the current version, the developed numerical code consists of a set of executable modules written in the Wolfram Language\textsuperscript{\copyright} and combined into the PEK package, so named in memory of my grandfather, Pavel E.~Kotelnikov, who was tortured in the GULAG (a system of forced labor camps) by the NKVD (the People's Commissariat for Internal Affairs), whose methods are inherited by the current generation of security officers in my country. 
As a result of recent modernization, PEK has acquired the ability to use interpolated experimental profiles of plasma pressure, magnetic field and lateral wall shape in addition to the model profiles that have been used up to the present time. .


To reduce repetition, the traditional review of the contents of many articles on the topic of interest is omitted and the reader is referred to the publications cited above. In addition to these, it is worth mentioning the work of Kesner et al.\
\citep{
     Kesner1985NF_25_275,
     LiKesnerLane1985NF_25_907,
     LiKesnerLane1987NF_27_101,
     LiKesnerLoDestro1987NF_27_1259
} with which the results of this work will be compared. In particular, it is worth noting the statement of Li, Kesner and LoDestro in ~\citep{LiKesnerLoDestro1987NF_27_1259}. They wrote that in a proportional chamber, the walls of which are as close as possible to the surface of the plasma column, anisotropic plasma is always stable near the threshold of mirror instability. Our calculations only partially confirm this statement.


Also, in order to shorten the introductory part of the article, the formulation of the LoDestro equation and boundary conditions has been moved to the Appendix \ref{A1}. It has been supplemented by recent findings, which have made it possible to calculate some more integrals analytically. To understand the main part of the article, it is enough to explain that it deals with the Sturm-Liouville problem for the LoDestro equation in the formulation described in detail in sections 2 and 5 of Ref.~\citep{Kotelnikov+2023NF_63_066027}.

Section \ref{s03} describes the anisotropic pressure distribution model in a mirror trap under both transverse and oblique NBI.  The beta limits imposed by the thresholds of mirror and fire hose instabilities are also calculated. Sections \ref{s04}, \ref{s05}, and \ref{s06} step by step present and discuss the results of calculations of critical betas $\beta_{\text{cr1}}$ and $\beta_{\text{cr2}}$, first for the case when there are no other means of MHD stabilization, except for the lateral conducting wall, next for the case when only end MHD anchors are mounted, and finally  for the case when infinitely strong MHD anchors stabilizers are used together on combination with lateral wall, In addition, section \ref{s06} explains how PEK package simulates MHD anchors with different stability margins. 

\subsection*{List of notations}

\begin{tabbing}
$M\quad $ \= Mirror ratio at the neck of the magnetic plug \\
$R$ \> Mirror ratio at the turning point of the sloshing ions\\
$L$ \> Mirror ratio at the location of the limiting ring or conducting end wall simulating\\
    \> the MHD anchor\\
$\psi$ \> Normalized magnetic flux with $\psi=1$ at the lateral boundary of the plasma column\\
$z$ \> Normalized coordinate along the axes of the plasma column with $z=0$ in \\
    \> the midplane of the trap and $z=\pm1$ on the magnetic plugs\\
$r$ \> Normalized radial coordinate $r=0$ on the axes of the trap and $r=a$ on \\
    \> the lateral surface of the plasma column\\
$a$ \> Normalized radius of the plasma column as a function $z$\\
$a_{0}$ \> Value of $a(z)$ in the trap midplane \\
$r_{w}$ \> Radius of the conducting cylinder surrounding the plasma column \\
$r_{w0}$ \> Value of $r_{w}$ in the trap midplane \\
$\Lambda$ \> Dimensionless function of the ratio $r_{w}(z)/a(z)$, included in the LoDestro equation \\
$\Lambda_{0}$ \> Value of $\Lambda(z)$ in the trap midplane \\
$b_{v}$ \> Normalized vacuum magnetic field as a function of $z$ with $b_{v}=1$ at \\
\> turning points of sloshing ions and $b_{v}=1/R$ in the mirror trap midplane \\
$b$ \> Normalized true magnetic field with $b=1$ at the turning points of sloshing ions \\
$p_{\bot}$ \> Normalized transverse plasma pressure as a function of $\psi$ and $z$\\
$p_{\|}$ \> Normalized longitudinal plasma pressure as a function of $\psi$ and $z$\\
$p_{0}$ \> Value of $p_{\bot}$ on the axis of the plasma column in the midplane of the trap\\
$F$ \> Distribution function of sloshing ions as a function of energy $\varepsilon $ and magnetic moment $\mu$ \\
$f_{k}$ \> Dimensionless function $\psi$ associated with the radial pressure profile with $f_{k}=1$\\
\> on the axes of the plasma column and $f_{k}=1$ on the lateral surface \\
$k$ \> Index of function $f_{k}$, where $k=\infty $ corresponds to the stepwise pressure\\
    \> profile with a sharp boundary, and $k=1$ corresponds to a quasi-parabolic profile \\
$q$ \> Index of the axial profile of the vacuum magnetic field, where $q=2$ corresponds\\
    \> to a quasi-parabolic profile, and larger $q$s simulate a profile with shorter magnetic plugs \\
Lw  \> Lateral wall configuration without MHD end anchors\\
Cw  \> Combined wall configuration with lateral wall and MHD end anchors, simulated by  \\
    \> conducting end plates located at the magnetic plugs \\ 
Bw  \> Blind wall configuration similar Cw to with weaker end MHD  stabilizers imitated \\
    \> by the conducting end plates located  behind magnetic plugs \\
Rw  \> Ring wall configuration with annular limiters located beyond the turning points \\
    \> and before magnetic plugs imitated by a conducting end plate located in front of \\
    \> the magnetic plug \\
Pr \> Proportional shape of the wall of the conducting cylinder surrounding the plasma column \\
St \> Straightened shape of the wall of the conducting cylinder surrounding the plasma column \\
$\beta $ \> The beta parameter is defined as the maximum of the ratio $2p_{\bot}/b_{v}^{2}$ \\
$\beta_{\text{cr}1} $ \> The first critical beta value defining the upper boundary of the lower \\
\> stability zone against rigid ballooning modes due to MHD end anchors \\
$\beta_{\text{cr}2} $ \> Second critical beta value defining the lower boundary of the upper \\
\> stability zone against rigid ballooning modes due to the lateral wall \\
$\beta_{\text{mm}} $ \> Beta value at the upper boundary of the stability zone against mirror modes\\
$\beta_{\text{fh}} $ \> Beta value at the upper boundary of the stability zone against hose modes\\
$\theta $ \> Pitch angle \\
$\theta_{\text{inj}} $ \> Inclination angle of NBI
\end{tabbing}

\section{Magnetic field model}\label{s02}

In paraxial approximation, the magnetic flux $\psi$ is related to the distance $r$ from the axis $z$ of an axially symmetric mirror trap by the equation
    \begin{gather}
    \label{03:04}
    \frac{r^{2}}{2} = \int_{0}^{\psi} \frac{\dif{\psi}}{B}
    .
    \end{gather}
The paraxial approximation means that the radius of curvature of the magnetic field lines significantly exceeds both the plasma radius and the distance between the magnetic mirrors. The equivalent condition for the applicability of such an approximation is formulated as the smallness of the plasma column radius compared to the distance between the magnetic mirrors.


The true magnetic field $B$ weakened by the diamagnetic effect is related to the vacuum magnetic field $B_{v}$ by the transverse equilibrium equation (where the rationalized electromagnetic units are used, they are also known as Heaviside—Lorentz units):
    \begin{gather}
    \label{03:08}
    B^{2} + 2P_{\bot} = B_{v}^{2}
    ,
    \end{gather}
where $P_{\bot}$ is the transverse pressure. In this form, it is approximately true in the paraxial approximation, when the radius of the plasma column $a=a(z)$ is everywhere small compared to the distance between the magnetic mirrors. 

In what follows, dimensionless notation is used, such that $b=B/B_{R}$, $b_{v}=B_{v}/B_{R}$, with  $B_{R}$ being the magnetic field at the turning point of fast ions, where the plasma transverse and longitudinal pressures drop to zero, $P_{\perp}=P_{\|}=0$, as explained in the next section. Instead of $P_{\perp}$ and $P_{\|}$, their normalized version $p_{\bot}=P_{\bot}/B_{R}^{2}$, $p_{\|}=P_{\|}/B_{R}^{2}$ are used below. In normalized units, Eq.~\eqref{03:08} reads
    \begin{gather}
    \label{03:09}
    b^{2} + 2p_{\bot} = b_{v}^{2}
    .
    \end{gather}

The  radial coordinate $r$ is normalized in such a way that $\psi=1$ at the plasma lateral boundary $r=a(z)$. The three-parameter function
    \begin{equation}
    \label{03:11}
    b_{v}(z)= \left[
        1 + (M-1)\sin^{q}(\pi z/2)
    \right]/R
    ,
    \end{equation}
earlier used by \citet{Kotelnikov+2022NF_62_096025, Kotelnikov+2023NF_63_066027,  ZengKotelnikov2024PPCF_66_075020}, approximates the axial profile of the vacuum magnetic field. It assume normalization of both the true magnetic field $B$ and the vacuum magnetic field $B_{v}$ by their common value at the tuning point $B_{R}$ of fast ions. In addition, it assumes that the magnetic mirrors are located in the $z=\pm1$ planes, which means that longitudinal coordinate $z$ is normalized by the distance between the midplane and the mirror plug throat. The repeated choice of the function $b_{v}(z)=B_{v}(z)/B_{R}$ in the form \eqref{03:11} is due to the desire to simplify the comparison of new results with previous publications.

\begin{figure}
  \centering
  \includegraphics[width=\linewidth]{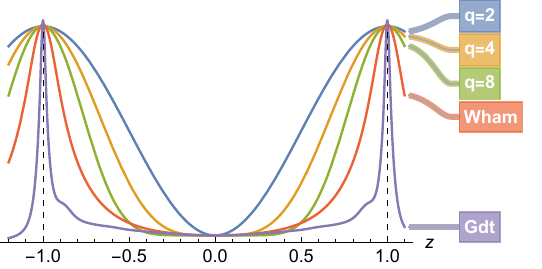}
  \caption{
    Axial profile of the vacuum magnetic field \eqref{03:11} for mirror ratio $M=32$ and three values of the index $q$ indicated on the graphs as compared to magnetic field in GDT \citep{Bagryansky+1990PPCNFR_2_655} and WHAM \citep{Endrizzi+2023JPP_89_975890501} devices.
  }\label{fig:Bv_vs_z_q}
\end{figure}
Parameter $q$ in Eq.~\eqref{03:11} determines the steepness and width of the magnetic mirrors: larger $q$ corresponds to the smaller fraction of the mirror plug in the total length of the trap, as shown in Fig.~\ref{fig:Bv_vs_z_q}.


Parameter $R=B_{R}/\min(B_{v})$ has the meaning of the mirror ratio between the turning point $B=B_{v}=B_{R}$, where the fast ions pressure drops to zero, and the minimum value $\min(B_{v})$ of the magnetic vacuum field in the middle plane of the trap. 


Parameter $M=\max(B_{v})/\min(B_{v})=B_{v}(\pm1)/B_{v}(0)$ is equal to the mirror ratio in the traditional sense.

\section{Anisotropic pressure models}\label{s03}

In publications on the stability of rigid ballooning mode, two analytic models of anisotropic pressure have previously been proposed. \emph{Kesner} in his paper \citep{Kesner1985NF_25_275}
indicates that in the first model the transverse pressure $p_{\bot}$ in a nonuniform magnetic field $b\leq 1$ varies according to the formula
    \begin{equation}
    \label{003:01}
    p_{\bot} = p(\psi) \left(
        1-b^{2}
    \right),
    \end{equation}
while in the second model
    \begin{equation}
    \label{003:02}
    p_{\bot} = p(\psi)\, b^{2}\left(
        1-b
    \right)^{n-1}
    ,
    \end{equation}
where $p(\psi)$ is a function of magnetic flux $\psi$, $b=1$ is the normalized magnetic field at a ``turning point'' where the fast ion pressure drops to zero, and $n\geq2$ is an integer index. Both models assume that in the region $b>1$ the plasma pressure is negligible, $p_{\perp}=0$, but there is a relatively cold plasma with low pressure that extends up to the magnetic field peak $b_{\max}=\max(B_{v})/\min(B_{v})=M/R$ in a magnetic plug. Cold plasma provides electrical contact with perfectly conducting end plates which imitate the end MHD anchors. 


The first model \eqref{003:01} was previously applied to the case of transverse injection of neutral atom beams (transverse NBI) \citep{Kotelnikov+2023NF_63_066027}. It describes the pressure distribution in an anisotropic plasma with a pressure peak near the magnetic field minimum at the midplane of the central cell of a mirror trap. The second model \eqref{003:02} is suitable for the case of an oblique NBI, when the pressure peak is located between the middle plane of the trap and the magnetic mirror. It is this model with indices $n=2$ and $n=3$ that is used in this article.

Kinetic theory  proves that if the transverse pressure $p_{\bot}$ is given as a function of $b$, then the longitudinal pressure $p_{\|}$ is uniquely determined using the parallel equilibrium equation with properly posed boundary condition \citep{Newcomb1981JPP_26_529}. The latter can be rewritten in terms of the partial derivative with respect to $b$ for a constant magnetic flux $\psi$ as
    \begin{equation}
    \label{003:03}
    p_{\bot} = - b^{2}\parder{}{b}\frac{p_{\|}}{b}
    .
    \end{equation}
It would be wrong therefore to choose at random a couple of functions for transverse and longitudinal pressures, since they must correspond to a distribution function, which is a solution to the kinetic equation. In section \ref{s3.ff} the distribution functions that lead to Eqs.~\eqref{003:01} and~\eqref{003:02} are restored under some dedicated assumptions.

Another key result of the kinetic theory is the assertion that the function $p_{\bot}/B^{2}$ always decreases as $B$ increases, i.e. \citep[Eq.~(37)]{Newcomb1981JPP_26_529}
    \begin{equation}
    \label{003:04}
    \parder{}{b}\frac{p_{\bot}}{b^{2}}
    \leqslant
    0
    .
    \end{equation}
Obviously, both models of anisotropic pressure satisfy the condition \eqref{003:04}. Note however that the inequality \eqref{003:04} was proved under the assumption that the class of distribution of fast ions $F(\varepsilon ,\mu)$ as function of energy $\varepsilon $ and magnetic moment $\mu$, is narrowed by the condition \citep[Eq.~(A6)]{Newcomb1981JPP_26_529}
\begin{gather}
    \label{003:05}
    \parder{F}{\varepsilon }\leq 0
\end{gather}
for all $\varepsilon $ and $\mu$, so that $F$ is everywhere monotone-decreasing in $\varepsilon $. It can be shown, in fact, that monotonicity in $\varepsilon $ is required as a condition for local variational stability in the sense of the Grad energy criterion \citep{Grad1967PF_10_137}. It also underlies the derivation of the Kruskal-Oberman energy principle \citep{KruskalOberman1958PF_1_275} and is considered by some authors as a necessary requirement for the correct formulation of the energy principle \citep{Stupakov1987Binp97-014eng}. Refusal of the condition \eqref{003:05} leads, in particular, to the appearance of new instabilities associated with resonances between longitudinal vibrations of ions and the wave \citep{SkovorodinZaytsevBeklemishev2013PoP_20_102123}.

%
%
%
%

It is known that anisotropic plasma can be subject to mirror (sometimes called diamagnetic) and fire hose instability (sometimes called garden hose instability). According to the fluid Chew‐Goldberger‐Low theory, stability of fire hose-type disturbances requires the inequality
    \begin{equation}
    \label{003:06}
    p_{\|}
    \leqslant
    p_{\bot}+b^{2}
    \end{equation}
to be hold for any $b$
\citep{
    Kulsrud1983OFP1eng,
    Ilgisonis1993PhysFluidsB_5_2387%
}.
Stability against the mirror mode implies that \citep{RudakovSagdeev1961DANUSSR_138_581, Thompson1964Introduction, Newcomb1981JPP_26_529, SouthwoodKivelson1993JGR_98_9181,  Ilgisonis1993PF_5_2387}
    \begin{equation}
    \label{003:07}
    \parder{}{b}
    \left(
        p_{\bot}+\frac{b^{2}}{2}
    \right)
    >
    0.
    \end{equation}
Simple, intuitively finger-tight, ways to derive criteria \eqref{003:06} and \eqref{003:07} are collected as the student problems after lecture 25 in the author's 2nd volume of ``Lectures on plasma physics'' \citep{Kotelnikov2022V2e}. All specified criteria of stability are satisfied for the pressure model \eqref{003:01}, which, recall, corresponds to the case of transverse NBI. For the oblique NBI model \eqref{003:02}, these conditions impose restrictions on the value of beta, which are specified below.

%

The fire hose instability is essentially a modification of the Alfven wave in an anisotropic plasma. In a homogeneous plasma, its increment is proportional to the longitudinal wave number $k_{\|}$. In the low-frequency limit, hose perturbations are strongly elongated along magnetic field lines and represent bending or torsional vibrations of magnetic flux tubes. Inhomogeneity of the magnetic field changes the threshold for the fire hose instability \citep{Mirnov1986diss(en)}.

In a homogeneous magnetic field, the increment of mirror instability is also proportional to the longitudinal wave number $k_{\|}$ and, oddly enough, inversely proportional to the number of resonant ions with zero longitudinal velocity \citep{SouthwoodKivelson1993JGR_98_9181, Pokhotelov+2002JGeoResSpacePhys_10_1312}. Near the threshold, mirror disturbances, like the flute ones, are strongly elongated along magnetic field lines.

The criteria \eqref{003:06} and \eqref{003:07} should be interpreted as formal and reference. For the pressure models described below, they first break down locally near two spatially separated points on the axis of the plasma column. It has not yet been studied whether the lateral conductive wall can influence the behavior of the localized mirror and fire hose disturbances. 

In the PEK's internal classification, the isotropic plasma variant is designated `A0', the anisotropic pressure \eqref{003:01}, which is formed under transverse NBI, is designated `A1'. The oblique NBI modeled by Eq.~\eqref{003:02} with an arbitrary index $n\geq2$ is denoted `A\emph{n}'. The two cases $n=2$ and $n=3$ discussed in detail in this article are abbreviated `A2' and `A3'.

Method of the MHD stabilization is specified by an additional abbreviation. Variants with conducting lateral wall stabilization are marked with the shortcut `Lw' (for Lateral wall), and variants with joint lateral wall and conducting end plates installed in the throat of the magnetic plug are marked with the shortcut `Cw' (for  Combined wall). The design of the lateral conducting wall in the form of a proportional chamber is labeled `Pr' (Proportional). Thus, the label `A3-LwPr' in a figure caption should be deciphered as the variant of plasma with anisotropic pressure of the `A3' type in a proportional conducting chamber without the end MHD anchor. Continuing the line of Ref.~\citep{ZengKotelnikov2024PPCF_66_075020},  this paper compares, when it is suitable, the stabilizing effect of proportional chamber with a straightened chamber, which is assigned the abbreviation `St' (Straightened).

In what follows, parameter beta is defined as the maximum of the ratio $2p_{\bot}/b_{v}^{2}$,
    \begin{gather}
    \label{003:08}
    \beta=\max(2p_{\bot}/b_{v}^{2})
    .
    \end{gather}
In case if the radial pressure profile is peaked at the trap axis, and the inequality \eqref{003:04} holds, the maximum is reached on the trap axis (where $\psi=0$) at the vacuum field minimum (where $\min(B_{v})=1/R$). Combining definition \eqref{003:08} and dimensionless version \eqref{03:09} of Eq.~\eqref{03:08} readily yields true magnetic field at the minimum
    \begin{gather}
    \label{003:09}
    b_{\min} = \sqrt{1-\beta }/R
    .
    \end{gather}

As shown in Refs.~\citep{Kotelnikov+2022NF_62_096025, Kotelnikov+2023NF_63_066027}, there can exist from none to two stable zones on the stability maps depending on the availability of lateral wall and MHD anchors. Using notations introduced in Ref.~\citep{ZengKotelnikov2024PPCF_66_075020}, one can specify than the Lw configuration can provide an upper stability zone at a sufficiently large beta that exceeds second critical value, $\beta > \beta_{\text{cr}2}$. In the Cw, Bw (Blind Wall) and Rw (Ring Wall) configurations lower zone $\beta < \beta_{\text{cr}1}$ can appear in addition to the upper one. These two zones can merge when the vacuum gap between the plasma column and the inner surface of the conducting shell is sufficiently narrow.

\subsection{Pressure model A1}\label{s3.A1}

The A1 pressure model is described in detail in Ref.~\citep{ZengKotelnikov2024PPCF_66_075020}. It is intended for modeling transverse NBI. To facilitate comparison with other pressure models, the main properties of the A1 model are briefly summarized below, starting with the definition
    \begin{gather}
    \label{003.1:01}
    \begin{aligned}
    p_{\bot} &= p_{0} f_k(\psi )\, \left(1-b^{2}\right)
    ,\\
    p_{\|} &= p_{0} f_k(\psi )\, \left(1-b\right)^2
    .
    \end{aligned}
    \end{gather}
The dimensionless function $f_{k}(\psi)$ defines the radial pressure profile and is normalized so that $f_{k}(0)=1$. It was chosen by \citet{Kotelnikov+2022NF_62_096025} so that the integrals over $\psi$ in the coefficients of the LoDestro equation could be calculated analytically at least for the case of isotropic plasma. For integer values of index $k$,
    \begin{equation}
    \label{003.1:02}
    f_{k}(\psi) =
    \begin{cases}
      1 - \psi^{k}, & \mbox{if } 0\leq \psi \leq 1 \\
      0, & \mbox{otherwise}
    \end{cases}
    \end{equation}
and for $k=\infty$
    \begin{equation}
    \label{003.1:03}
    f_{\infty}(\psi) = H(1-\psi)
    ,
    \end{equation}
where $H(x)=0$ for $x<0$ and $H(x)=1$ for $x>0$. In the case of oblique NBI, the integrals over $\psi$ in the coefficients of the LoDestro equations can calculated only numerically, and, therefore there is no needs to restrict class of functions $f_{k}(\psi)$, except for the comparison of the new and earlier results. Function $f_{1}$ describes the most smooth radial pressure profile. Maximal index $k=\infty $ corresponds to a stepwise profile  with sharp boundary.

Parameter $p_{0}$ as defined in this paper has no dedicated meaning. As explained by \cite{Kotelnikov+2023NF_63_066027}, scanning for the solution of the Sturm-Liouville problem over the parameter $p_0$ makes the numerical code simpler than scanning over $\beta$. Parameter $p_0$ can be redefined to prescibe it the value of the transverse pressure at $r=z=0$. However, such a choice would make many equations more complicated.

Equation of transverse equilibrium \eqref{03:09} is readily resolved regarding true magnetic field as
    \begin{gather}
    \label{003.1:03}
    b(z,\psi)
    = 
    \sqrt{
        \frac{b_{v}^{2}(z)-2p_{0}f_{k}(\psi)}{1-2p_{0}f_{k}(\psi)}
    }
    \,.
    \end{gather}
Plasma's beta defined by Eq.~\eqref{003:08} is related to parameter $p_{0}$ in Eq.~\eqref{003.1:01} by 
    \begin{gather}
    \label{003.1:05}
    \beta
    =
    \frac{2 \left(R^2-1\right)p_{0}}{1 - 2p_{0}}
    .
    \end{gather}
Parameter $p_{0}$ can vary within the range 
\begin{gather}
0<p_{0}<1/2R^{2},
\end{gather}
and $\beta \to 1$ for $p_{0}\to 1/2R^{2} $. For $\beta >1$, transverse equilibrium near the median plane of the trap is impossible, at least in paraxial approximation and within the MHD approach.\footnote{Equilibrium with $\beta>1$ is still possible in systems where ion Larmour radius is large in a certain sense, see \citep{Kotelnikov2020PPCF_62_075002, KurshakovTimofeev2023PoP_30_092513, Timofeev+2024PoP_31_08512, KhristiBeklemishev2024Srxiv_2408.06792v1}.}

\begin{figure*}
  \centering
  \includegraphics[width=0.3\linewidth]{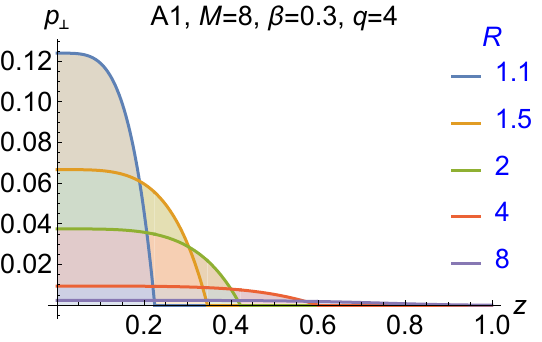}
  \hfil
  \includegraphics[width=0.3\linewidth]{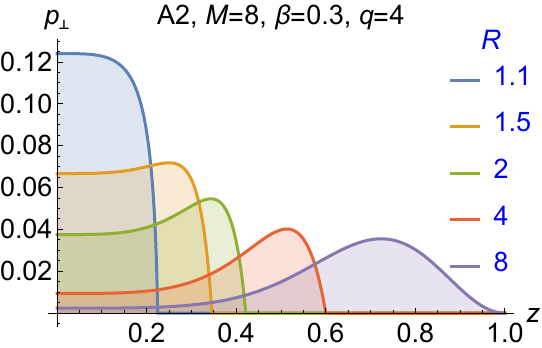}
  \hfil
  \includegraphics[width=0.3\linewidth]{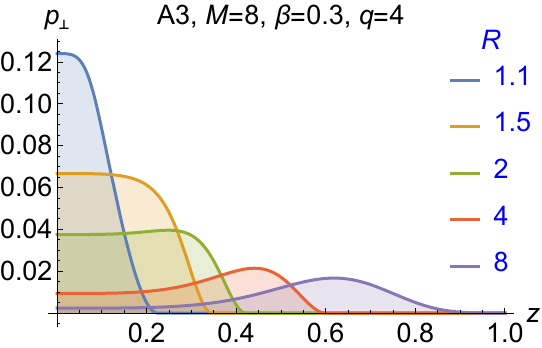}
  \caption{
    Axial profiles of the transverse pressure in models A1 (left), A2 (center) and A3 (right) described respectively by Eqs.~\eqref{003.1:01}, \eqref{003.2:01}, \eqref{003.3:01} in magnetic field \eqref{03:11} at the axis of the trap for mirror ratio $M=8$, index $q=4$ and various mirror ratios $R$ at the turning points where pressure of hot plasma component drops to zero. 
    The value $\beta=0.3$ for this figure is chosen so as not to exceed the threshold for excitation of mirror and fire hose instabilities for all values of the parameter $R$ indicated in the figure legend.
  }\label{fig:GM2-A123-Pt_vs_R-M8_q4}
\end{figure*}

The left graph in Fig.~\ref{fig:GM2-A123-Pt_vs_R-M8_q4} shows axial profile of the transverse pressure for the model A1 in magnetic field \eqref{03:11}. The pressure peak in this model is located at the midplane of the trap and narrows as the local mirror ratio $R$ at the turning point of fast ions decreases. 

It seems intuitively obvious that the degree of anisotropy also increases as $R$ decreases. However, it is not so easy to give a completely satisfactory definition of the degree of anisotropy suitable for all pressure models to be considered in this paper. If the degree of anisotropy would be related to the pressure in the minimum magnetic field as 
\begin{gather}
\label{003.1:06}
A_{1} = \frac{p_{\bot}-p_{\|}}{p_{\bot}+p_{\|}} 
,
\end{gather}
then 
\begin{gather}
\label{003.1:07}
A_{1} 
=
{\sqrt{1-\beta}}/{R}
\end{gather}
which it is indeed inversely proportional to the parameter $R$.

It can be proved that the A1 pressure model is stable against the mirror and fire hose modes. For example, putting first of Eq.~\eqref{003.1:01} into Eq.~\eqref{003:07} yields the inequality $b\,(1-2p_{0})>0$, which is incompatible with condition $0<p_{0}<1/2R^{2}$ within the range $0<b<1$.

\subsection{Pressure model A2}\label{s3.A2}

The  model \eqref{003:02} with index $n=2$ reduces to
    \begin{gather}
    \label{003.2:01}
    \begin{aligned}
    p_{\bot} &= p_{0} f_k(\psi )\, b^2\left(1-b\right)
    ,\\
    p_{\|} &= \frac{1}{2} p_{0} f_k(\psi )\, b\left(1-b\right)^2
    \end{aligned}
    \end{gather}
in the region $b<1$.

In case of pressure model A1 \eqref{003:01}, Eq.~\eqref{03:09} reduces to a second-order polynomial equation with respect to $b$. For the model A2 same  equation is of the third order, and the model A3 at $n=3$ yields a fourth order algebraic equation. As is well known, the solution to such equations can always be expressed in terms of radicals, which saves us from the need to solve these expressions numerically. However, the resulting formulas for a polynomial of the third (and even more so, fourth) order are extremely cumbersome. Therefore, the result of solving Eq.~\eqref{03:09} is written below in a more compact form, which is introduced by Wolfram \emph{Mathematica}$^{\copyright}$:
    \begin{gather}
    \label{003.2:02}
    b
    =
    \Root\left[
        2 {\#}^3 p+{\#}^2 (-2 p-1) +b_v^2
        \,\&,\,2
    \right]
    .
    \end{gather}
Literally, it means the second root of the cubic equation
    \begin{gather*}
    2 {b}^3 p+{b}^2 (-2 p-1) +b_v^2
    =0
    \end{gather*}
represented as a ``pure function'' by the ampersand $\&$ in the first argument of the $\Root$ built-in utility. Parameter $p$ in this formula stands for $p_{0}f_{k}(\psi)$. The $\Root$ utility arranges the roots of the polynomials in an order known only to it, but in such a way that the real roots come first.

Relative plasma pressure, i.e.\ parameter beta, is defined by Eq.~\eqref{003:08} as the maximum of the ratio $2p_{\bot}/b_{v}^{2}$. According to Eq.~\eqref{003:04}, this ratio is a decreasing function of $b$ so that the maximum is reached on the axis of the trap (where $\psi=0$) at the minimum of the vacuum field (where $b_{v}=1/R$). Hence
    \begin{gather}
    \label{003.2:06}
    p_{0}
    =
    \frac{\beta}{2 (1-\beta) \left(1-\sqrt{1-\beta }/R\right)}
    .
    \end{gather}
Inversion of Eq.~\eqref{003.2:06} yields
    \begin{multline}
    \label{003.2:08}
    \beta
    =
    \Root
    \left[
        4 {\#}^3 p_{0}^2
        +
        {\#}^2
        \left(
            4 p_{0}^2R^2-12 p_{0}^2+4 p_{0} R^2
            +
            R^2
        \right)
    \right.
    \\
    \left.
        -
        {\#}
        \left(8 p_{0}^2 R^2-12 p_{0}^2+4 p_{0} R^2\right)
        -
        4 p_{0}^2 R^2+4 p_{0}^2
        \,\&,2
    \right]
    .
    \end{multline}

As follows from Eq.~\eqref{003.2:06}, parameter $p_{0}$ tends to infinity as $\beta\to 1$. In fact, the mirror mode stability condition \eqref{003:07} imposes a more stringent condition 
\begin{gather}
    \label{003.2:08a}
    0<p_{0}<1. 
\end{gather}
If $p_{0}>1$, there is no continuous solution to Eq.~\eqref{03:09}. This can be verified by plotting the left side of the equation as a function of $b$. For $p_{0}>1$, the maximum of this function on the interval $0<b<1$ exceeds $1$, while the left side of the equation is less than $1$. This means that paraxial equilibrium is impossible above the mirror instability threshold. Near this threshold, amusing phenomena such as equilibrium hysteresis appear. Plasma can reside in two different states, between which a transition is possible by relieving excess pressure \citep{Kotelnikov2011FST_59_47}. Signs of such a transition were found in experiments at the GDT facility \citep{Kotelnikov+2010PhysRevE_81_067402}. Similar structures are observed in the solar wind, see e.g.\ \citep{Winterhalter+1994JGeoPhysRes_99_23371, Pantellini1998JGeoResSpacePhys_103_4789, Jiang+2022AAS_935_169}.


In practice, it turned out that the upper limit of the interval \eqref{003.2:08a} is sometimes inaccessible to the shooting method when solving the LoDestro equation \eqref{A2:01}. The solution of this equation with initial conditions \eqref{A2:14} and \eqref{A2:15} on the left boundary $z=0$ of the domain of definition of the Sturm-Liouville problem (see Appendix \ref{A1}) sometimes went to infinity, before reaching the right boundary $z=z_{E}$. Most often this happened if the parameter $p_{0}$ was close enough to $p_{0}=1$.

The threshold value of parameter $p_{0}$, above which the mirror instability is excited, will be further written as a function
    \begin{gather}
    \label{003.2:09}
    p_{\text{mm}}(R) = 1,
    \end{gather}
where the subscript `mm' stands for ``mirror mode''. The beta limit is obtained from Eq.~\eqref{003.2:08} by substituting $p_{0}=1$:
    \begin{gather}
    \label{003.2:11}
    \beta_{\text{mm}}(R) =
    \Root
    \left[
        4 {\#}^3
        +
        {\#}^2 \left(9 R^2-12\right)
        +
        {\#} \left(12-12 R^2\right)+4 R^2-4
        \,\&,2
    \right]
    .
    \end{gather}
\begin{figure*}
  \centering
  \includegraphics[width=0.48\linewidth]{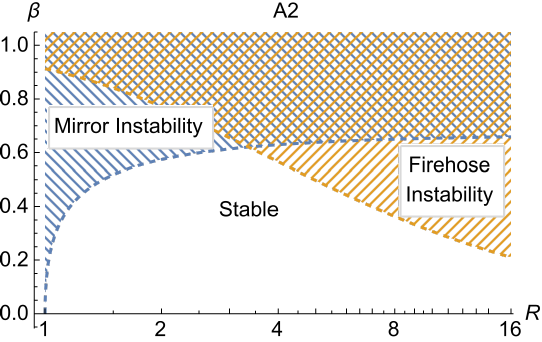}
  \hfil
  \includegraphics[width=0.48\linewidth]{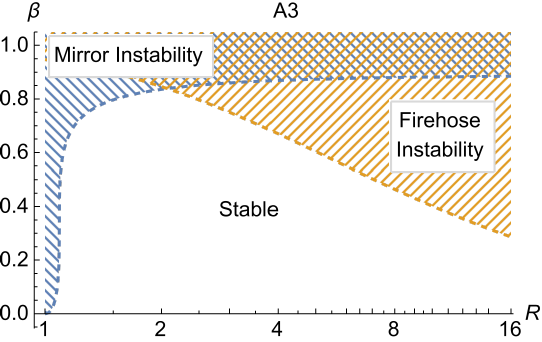}
  \caption{
    Threshold of mirror and fire hose instabilities in an anisotropic plasma with oblique NBI within the framework of models A2 (left) and A3 (right). Unstable areas are hatched. The same shading scheme is used below in the stability maps without further reminder. Mirror instability (blue hatching) dominates at low $R$ while fire hose mode (orange hatching) dominates at large $R$.
  }
  \label{fig:StableZone-A2A3-beta_vs_R}
\end{figure*}
The fire hose instability threshold is described by more cumbersome formulas:
    \begin{gather}
    \label{003.2:12}
    p_{\text{fh}}(R)
    =
    \Root
    \left[
        {\#}^3 \left( 4 R^4 + 4 R^2\right)
        +
        {\#}^2 \left(R^4-30 R^2-27\right)
        -
        24 {\#} R^2
        -4 R^2
        \,\&,j
    \right],
    \end{gather}
    \begin{gather}
    \label{003.2:14}
    \beta_{\text{fh}}(R)
    =
    \Root
    \left[
        {\#}^3
        +
        {\#}^2\left(R^2-9\right)
        +
        24 {\#}
        -
        16
    \,\&,j
    \right]
    ,
    \end{gather}
where $j=1$ if $1<R<\sqrt{9+6 \sqrt{3}}$ and $j=3$ if $R>\sqrt{9+6 \sqrt{3}}$; the subscript `fm' is a shortcut for ``fire hose mode''.

%
The stability zone bounded by the conditions $\beta<\beta_{\text{mm}}(R)$ and $\beta<\beta_{\text{fh}}(R)$ is shown on the left in Fig.~\ref{fig:StableZone-A2A3-beta_vs_R}. Its width shrinks to zero both in the $R\to1$ limit and in the $R\to\infty$ limit. The maximum width corresponds to the point of intersection of the curves $\beta=\beta_{\text{mm}}(R)$ and $\beta=\beta_{\text{fh}}(R)$, where $\beta = \num{ 0.620204}$, $R=\num{3.35839}$. At the same time, it is worth noting that at the maximum of these functions are equal to $\beta_{\text{mm}}(\infty )=2/3$ and $\beta_{\text{fh}}(1)=\num{0.9126219746158469}$.

Limitation of the stability zone in an anisotropic plasma by the threshold of mirror instability was taken into account earlier by Kesner et al. \citep{Kesner1985NF_25_275, LiKesnerLoDestro1987NF_27_1259}, but the fire hose instability did not attract the attention of these authors.

Exceeding the threshold of mirror instability within the framework of the applicability of the LoDestro equation is impossible, since this is equivalent to a violation of the equilibrium configuration of the plasma in the paraxial approximation, which is indirectly confirmed by experiments at the Gas-Dynamic Trap \citep{Kotelnikov+2010PhysRevE_81_067402}. The PEK code refused to continue computing if $p_{0}>p_{\text{mm}}$. However, with a few tweaks, it works above the fire hose instability threshold. In some cases, it finds a solution in the interval $p_{\text{fh}}(R)<p_{0}<p_{\text{mm}}(R)$.


There are a number of theoretical papers \citep{Kotelnikov+2010PhysRevE_81_067402, Kotelnikov2011FST_59_47, Beklemishev2016PoP_23_082506} that predict the stabilization of mirror instability in a nonuniform magnetic field. According to these works, when the mirror instability threshold is exceeded, a region appears in the plasma near the trap axis, where something like a magnetic field jump is formed, and nonparaxiality effects smooth out this jump. It is not yet clear at the moment whether the LoDestro equation can be modified to extend its range of applicability to equilibria with narrow non-paraxial jumps inside the plasma column.

The condition \eqref{003:06} guarantees the stability of small-scale perturbations of the fire hose type, which, if the parameter $p_{0}$ slightly exceeds $p_{\text{fh}}(R)$, can become unstable near the axis of the plasma column. It is expected that such perturbations are not critical for the rigid ballooning mode.

\begin{figure*}
  \centering
  \includegraphics[width=0.3\linewidth]{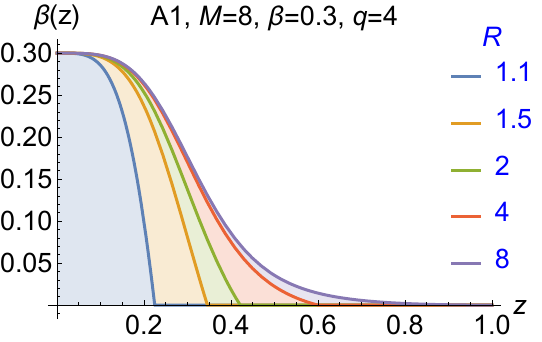}
  \hfil
  \includegraphics[width=0.3\linewidth]{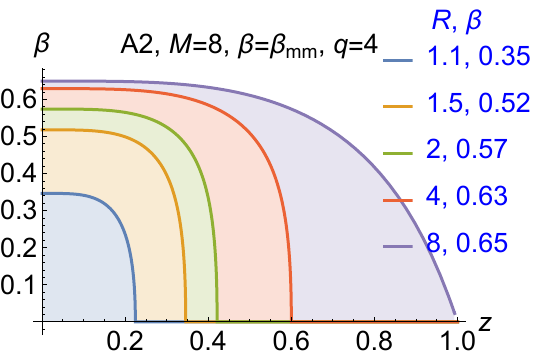}
  \hfil
  \includegraphics[width=0.3\linewidth]{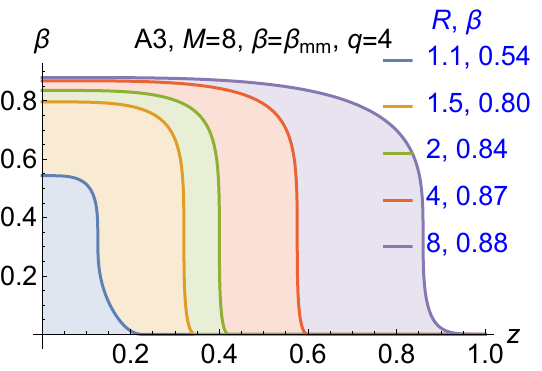}
  \caption{
    Axial profile of the local beta $\beta(z) = 2p_{\bot}/B_{v}^{2}$ in the magnetic field \eqref{03:11} for mirror ratio $M=8$, index $ q=4$ and various mirror ratios $R$ at the turning points where pressure of hot plasma component drops to zero. The values of the $\beta $ parameter for each $R$ value indicated on the graphs are chosen to be equal to the smallest of the two stability thresholds for mirror and fire hose instability.
    }
    \label{fig:GM2-A123-beta_vs_R-M8_q2}    
\end{figure*}
Figure \ref{fig:GM2-A123-beta_vs_R-M8_q2} confirms the above statement that the relation $\beta(z) = 2p_{\bot}/B_{v}^{2}$, which has the meaning of local beta, has a global maximum in the median plane of the trap. Moreover, $\beta(z)$ decreases monotonically in a monotonically increasing magnetic field. However, kinetic calculations of the distribution function of sloshing ions using various numerical codes, in particular the DOL code \citep{YurovPrikhodkoTsidulko2016PPR_42_210}, show that if the condition \eqref{003:05} is violated, the $\beta(z)$ dependence may turn out to be non-monotonic and a local peak appears on the $\beta(z)$ graph near fast ion turning point.

\subsection{Pressure model A3}\label{s3.A3}

In the model A3, the pressure functions have the form
    \begin{gather}
    \label{003.3:01}
    \begin{aligned}
    p_{\bot} &= 
    p_{0} f_k(\psi )\, b^2\left(1-b\right)^{2}
    ,\\
    p_{\|} &=
    \tfrac{1}{3}\,
    p_{0} f_k(\psi )\, b\left(1-b\right)^3
    .
    \end{aligned}
    \end{gather}
In at least one respect, the A3 model is more realistic than other anisotropic pressure models, discussed so far. The point is that in models A1 and A2 the derivative $\tparder{p_{\bot}}{b}$ suffers a discontinuity at $b=1$. It should be expected that the actual pressure distribution in a mirror trap should not have such a discontinuity, since it will inevitably be smoothed out by Coulomb collisions of plasma particles. At the discontinuity point, the derivative $\tparder{b}{z}$ tends to infinity, as on the threshold of mirror instability. In the A3 model, the derivative $\tparder{b}{z}$ also tends to infinity as it approaches the mirror instability threshold, but this happens at some distance from the turning point $b=1$, namely, at $b=3/4 $. 

Comparing subfigures in Fig.~\ref{fig:GM2-A123-Pt_vs_R-M8_q4}, from the left to the right, which show the axial transverse pressure profiles for models A1, A2, and A3, respectively, one can see that in the latter subfigure the pressure profile is smoother near the turning point. Another difference is that for the same values of the parameter $R$, the maximum pressure in the right subfigure is shifted closer to the center of the trap compared to the central subfigure.

It was noted in Refs.~\citep{Kotelnikov+2023NF_63_066027, ZengKotelnikov2024PPCF_66_075020} that parameter $R$ characterizes the degree of plasma anisotropy in the A1 model. According to Eq.~\eqref{003.1:07}, the degree of anisotropy is higher for smaller $R$. In the case of oblique NBI \eqref{003:02} this connection does not seem obvious. Kesner \citep{Kesner1985NF_25_275} defines the degree of anisotropy as the ratio of the transverse pressure to the longitudinal pressure at the magnetic field minimum:
    \begin{gather}
    \label{003.3:02}
    A_{\text{K}}
    =
    {p_{\bot}}/{p_{\|}}
    \qquad 
    \text{ at }\ b=b_{\min}=\sqrt{1-\beta }/R.
    \end{gather}
Examining the quantity $p_{\bot}/p_{\|}=nb/(1-b)$ at an arbitrary $b$ reveals it monotonically increases up to infinity as it approaches the turning point $b=1$. The point $b=b_{\min}$ is distinguished only by the fact that the local beta maximum is located there. However, in the model \eqref{003:02} the transverse pressure peak is in the field $b=2/(n+1)$, where the ratio ${p_{\bot} }/{p_{\|}}=2n/(n-1)$ does not depend on $R$. This and other similar facts indicate that model \eqref{003:02} with any $n$  actually describes the pressure distribution with approximately fixed degree of anisotropy. 

In the next section \ref{s3.ff} it is shown that parameter $R$ is related to the angle of injection of neutral atoms. Note however that it is possible to invent such a definition of the degree of anisotropy that yields the same result as Eq.~\eqref{003.1:06} for the A1 model, namely
    \begin{gather*}
    A_{n}  = \frac{p_{\bot}}{p_{\bot}+n p_{\|}} 
    = \frac{\sqrt{1-\beta }}{R}
    .
    \end{gather*}

In the A3 model, the true magnetic field in the plasma is expressed in terms of the real root of a fourth-degree polynomial equation:
\begin{gather}
    \label{003.3:04}
    b
    =
    \Root\left[
        2 {\#}^4 p
        -
        4 {\#}^3 p
        +
        {\#}^2 (2 p+1)
        -b_{v}^2
        \,\&,2
    \right]
    .
\end{gather}
Parameter  $\beta $ is still defined by Eq.~\eqref{003:08}, but Eq.~\eqref{003.2:06} should be replaced by
\begin{gather}
    \label{003.3:05}
    p_{0}
    =
    \frac{\beta}{2 (1-\beta) \left(1-\sqrt{1-\beta }/R\right)^2}
    .
\end{gather}
The inverse expression is noticeably more complex than \eqref{003.2:08}:
\begin{multline}
    \label{003.3:06}
    \beta
    =
\Root\left[
    4 \#^4 p_{0}^2
    +
    \#^3
    \left(8 p_{0}^2 R^2-16 p_{0}^2-4 p_{0} R^2\right)
\right.
\\
\left.
    +
    \#^2\left(4 p_{0}^2 R^4-24 p_{0}^2 R^2+24 p_{0}^2+4 p_{0} R^4+8p_{0}    R^2+R^4\right)
\right.
\\
\left.
    +
    \# \left(-8 p_{0}^2 R^4+24 p_{0}^2 R^2-16 p_{0}^2-4 p_{0} R^4-4 p_{0} R^2\right)
\right.
\\
\left.
    +
    4 p_{0}^2 R^4-8 p_{0}^2 R^2+4 p_{0}^2\&\,,1
\right]
    .
\end{multline}

The threshold value of $p_{0}$, above which the mirror instability is excited, is now greater than prescribed by Eq.~\eqref{003.2:09}:
    \begin{gather}
    \label{003.3:07}
    p_{\text{mm}}(R) = 4
    .
    \end{gather}
The beta threshold is computed as a root of Eq.~\eqref{003.3:06} at $p=4$:
\begin{multline}
    \label{003.3:08}
    \beta_{\text{mm}}(R)
    =
    \Root\left[
        64 {\#}^4
        +
        {\#}^3 \left(112 R^2-256\right)
        +
        {\#}^2 \left(81 R^4-352 R^2+384\right)
    \right.
    \\
    \left.
        +
        {\#} \left(-144 R^4+368 R^2-256\right)
        +
        64 R^4-128 R^2+64
       \,\&,1
    \right]
    .
\end{multline}
The fire hose instability threshold is described by equally cumbersome formulas:
\begin{multline}
    \label{003.3:09}
    p_{\text{fh}}(R)
    =
    \Root\left[
        {\#}^3 \left(9R^6+18R^4+9R^2\right)
        +
    \right.
    \\
    \left.
        {\#}^2 \left(2R^6-150R^4-264R^2-128\right)
        +
        {\#} \left(96R^2-117R^4\right)
        -
        18R^4
        \,\&,j
    \right]
   ,
\end{multline}
where $j=1$ if
    $
    R
    <
    \sqrt{\frac{1}{2} \left(59+11\sqrt{33}\right)}
    $,
and $j=3$ if
    $
    R
    >
    \sqrt{\frac{1}{2} \left(59+11\sqrt{33}\right)}
    $.
But the fire hose threshold over beta is written almost as simply as \eqref{003.2:14}:
    \begin{gather}
    \label{003.3:16}
    \beta_{\text{fh}}(R)
    =
    \Root\left[
        4 {\#}^3+{\#}^2 \left(R^2-28\right)
        +60 {\#}-36
    \,\&,j
    \right]
    ,
    \end{gather}
where 
    $j=1$ if $1 R < R_{\ast}=\frac{1}{2}\Root\left[{\#}^4-236 {\#}^2-2048\,\&,2\right]$, 
and 
    $j=3$ if $  R > R_{\ast}$.

The stability zone bounded by the conditions $\beta<\beta_{\text{mm}}(R)$ and $\beta<\beta_{\text{fh}}(R)$ is shown in Fig.~\ref{fig:StableZone-A2A3-beta_vs_R} on the right. Its width shrinks to zero both in the $R\to1$ limit and in the $R\to\infty$ limit. The maximum width corresponds to the point of intersection of the curves $\beta=\beta_{\text{mm}}(R)$ and $\beta=\beta_{\text{fh}}(R)$, where $\beta=\num{0.838632}$, $R=\num{2.0706}$. At the same time, it is worth noting that the maxima of these functions are respectively $\beta_{\text{mm}}(\infty )=8/9$ and $\beta_{\text{fh}}(1)=\num{0.946833}$. Comparing left and right subfigures in Fig.~\ref{fig:StableZone-A2A3-beta_vs_R} one can see that in model A3 this zone is noticeably larger than in model A2. 
As can be assumed by looking at the graphs in Fig.~\ref{fig:AngleDistribution_n1238_vs_theta} in the next section, this fact is associated with the observation that the distribution function in model A3 is slightly closer to isotropic than in model A2.

\subsection{Angle distribution of fast ions}
\label{s3.ff}

For an arbitrary index $n$, the oblique injection model reads
    \begin{gather}
    \label{003.4:01}
    p_{\bot} =  p_{0} f_k(\psi )\,
    b^2\left(1-b\right)^{n-1}
    ,\\
    \label{003.4:02}
    p_{\|} =
    p_{0} f_k(\psi )\,
    \frac{b}{n}\left(1-b\right)^{n}
    ,\\
    \label{003.4:03}
    p_{0} = \frac{\beta }{
        2\left(1-\beta\right)
        \left(1-\sqrt{1-\beta}/R
    \right)^{n-1}}
    .
    \end{gather}
The threshold value of parameter $p_{0}$ with respect to the excitation of the mirror instability is given by
    \begin{gather}
    \label{003.4:04}
    p_{\text{mm}}(R) =
    \left(
        \frac{n-2}{n+1}
    \right)^{2-n}
    .
    \end{gather}
It is not possible to write formulas for $\beta_{\text{mm}}(R)$, $p_{\text{fh}}(R)$, $\beta_{\text{fh}}(R)$ in compact form.


Let's try to find an answer to the question, how realistic are the analytical models of anisotropic pressure? Is it possible to specify the distribution function of fast ions, which forms the pressure profiles that are given by Eqs.~\eqref{003:01} and~\eqref{003:02}?

It is known that if the distribution of fast ions is given by a function $F(\varepsilon,\mu)$ of two variables: the energy $\varepsilon $ and the magnetic moment $\mu$, then the dependence of the density $N$, transverse $p_{\bot}$ and longitudinal pressure $p_{\|}$ on the magnetic field can be calculated through double integrals of the distribution function over the variables $\varepsilon$ and $\mu$ with different weight factors:
    \begin{gather}
    \label{003.4:05}
    N(B)
    =
    \frac{2 \sqrt{2} \pi  B}{m^{3/2}}
    \int_{0}^{\infty}
    \left(\int_{\epsilon/B_{R}}^{\epsilon/B}
    \frac{F}{\sqrt{\epsilon -\mu B}}
    \,
    \dif\mu \right)
    \,
    \dif\epsilon
    ,\\
    \label{003.4:06}
    p_{\bot}(B)
    =
    \frac{2 \sqrt{2} \pi  B^{2}}{m^{3/2}}
    \int_{0}^{\infty}
    \left(\int_{\epsilon/B_{R}}^{\epsilon/B}
    \frac{\mu F}{\sqrt{\epsilon -\mu B}}
    \,
    \dif\mu \right)
    \,
    \dif\epsilon
    ,\\
    \label{003.4:07}
    p_{\|}(B)
    =
    \frac{4 \sqrt{2} \pi  B}{m^{3/2}}
    \int_{0}^{\infty}
    \left(\int_{\epsilon/B_{R}}^{\epsilon/B}
    \sqrt{\epsilon -\mu B}
    \,
    F\,
    \dif\mu \right)
    \,
    \dif\epsilon
    .
    \end{gather}
The inverse problem of recovering the distribution function $F$ depending on two variables from a given dependence of $p_{\bot}(B)$ on one variable obviously does not have a unique solution. The situation will not change if $p_{\|}(B)$ is added to $p_{\bot}(B)$, since these two functions are related by the longitudinal equilibrium equation. The function $N(B)$ is not yet known to us.

The situation changes radically if the distribution function can be rewritten in the form of $F=F(\varepsilon,\mu/\varepsilon)$, for example, $F=g(\varepsilon)\,F_{n}(\mu B_{R}/\varepsilon )$; in what follows, the subscript $n$ in the notation $F_{n}$ is related to the pressure model index A$n$. Then the integrals are separated. By replacing $\mu=\varepsilon\sin^{2}\theta/B_{\min}$, $B=B_{R}b$, $B_{R}=B_{\min}R$, $\sin^{2}\theta =X/R$ in Eq.~\eqref{003.4:06}, one obtains
    \begin{gather}
    \label{003.4:08}
    p_{\bot}(b)
    =
    \frac{2\sqrt{2}\pi b^{2}}{m^{3/2}}
    \left(
        \int_{0}^{\infty}
        \varepsilon^{3/2}g(\epsilon )
        \,
        d\epsilon
    \right)
    \left(
        \int_{1}^{1/b}
        \frac{X F_{n}(X)}{\sqrt{1 -b X}}
        \,
        dX
    \right)
    .
    \end{gather}
The integral in the first pair of brackets gives a constant independent of $b$. Omitting this and other constants leads to the Volterra integral equation of the first kind (see, e.g., \cite[chap.~10]{Polyanin+2008HandbookIntegralEquations}) for determining the function $F_{n}(X)$:
    \begin{gather}
    \label{003.4:11}
    (1-b)^{n-1}
    =
    \int_{1}^{1/b}
    \frac{X F_{n}(X)}{\sqrt{1-b X}}
    \,
    \dif{X}
    .
    \end{gather}
It is solved using the Laplace transform and the convolution theorem. To bring the last equation to the canonical formulation of the convolution theorem, one needs to make a few more changes of variables: $b=1/t$; $t=1+\tau$, $X=1+x$. It yields the equation
    \begin{gather}
    \label{003.4:12}
    h_{n}(\tau)
    =
    \frac{\tau^{n-1}}{\left(1+\tau\right)^{n-1/2}}
    =
    \int_{0}^{\tau}
    K(\tau-x)\,Q_{n}(x)
    \dif{x}
    ,
    \end{gather}
where $K(\tau)=1/\sqrt{\tau}$ is the kernel of the integral  equation, and $Q_{n}(x)=(x+1)F_{n}(x+1)$. By the convolution theorem, the Laplace image of the desired function $Q_{n}(x)$ is equal to the Laplace image of the function $h_{n}(\tau)$ on the left side of Eq.~\eqref{003.4:12} divided by the Laplace image of $K (\tau)$. After performing the inverse Laplace transform, it is possible to restore the function $F_{n}(X)$. The result takes unexpectedly simple form 
    \begin{gather}
    \label{003.4:14}
    F_{n}(X)
    =
    \frac{\Gamma(n)}{\sqrt{\pi}\Gamma(n-1/2)}
    \frac{(X-1)^{n-3/2}}{X^{n+1}}
    .
    \end{gather}
Recall that $X=R\sin^{2}\theta $, where $\theta$ is the pitch angle in the median plane. Function \eqref{003.4:14} passes through a maximum at $X=2(n+1)/5$. In terms of the pitch angle, the last relation means that
    \begin{gather}
    \label{003.4:15}
    \sin^{2}\theta = 2(n+1)/5R
    .
    \end{gather}
Thus, parameter $R$ can be interpreted as a measure of the slope of the NBI.

\begin{figure*}
  \centering
  \includegraphics[width=0.45\linewidth]{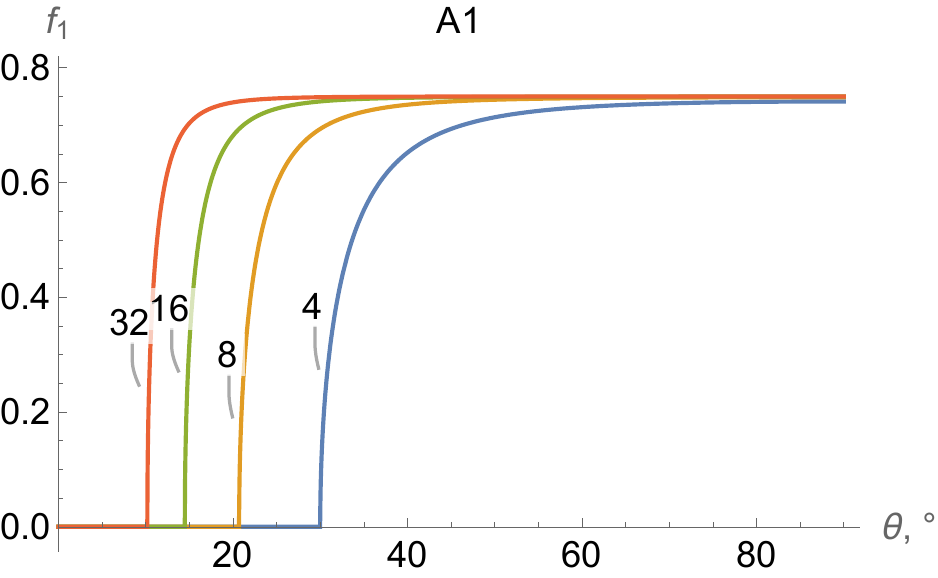}\hfil
  \includegraphics[width=0.45\linewidth]{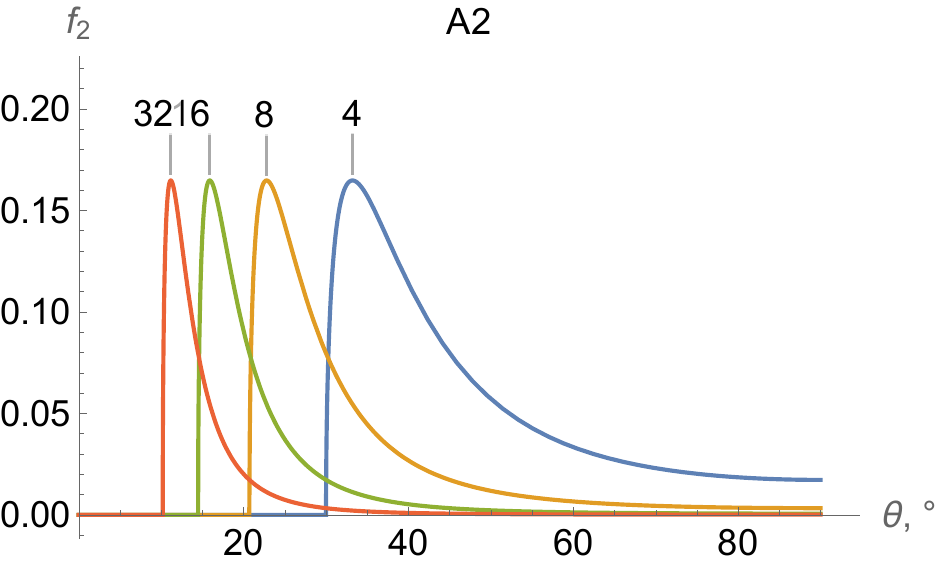}\\
  \includegraphics[width=0.45\linewidth]{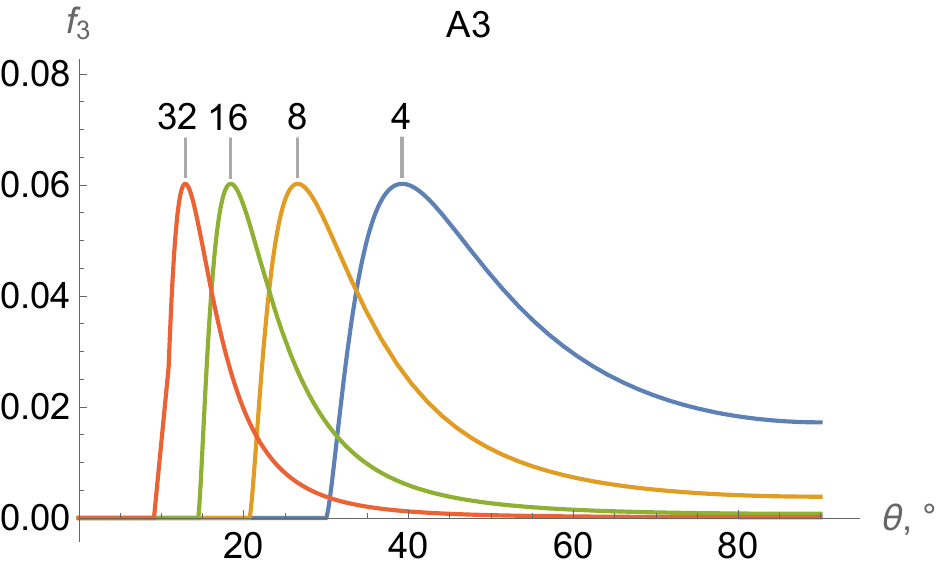}\hfil
  \includegraphics[width=0.45\linewidth]{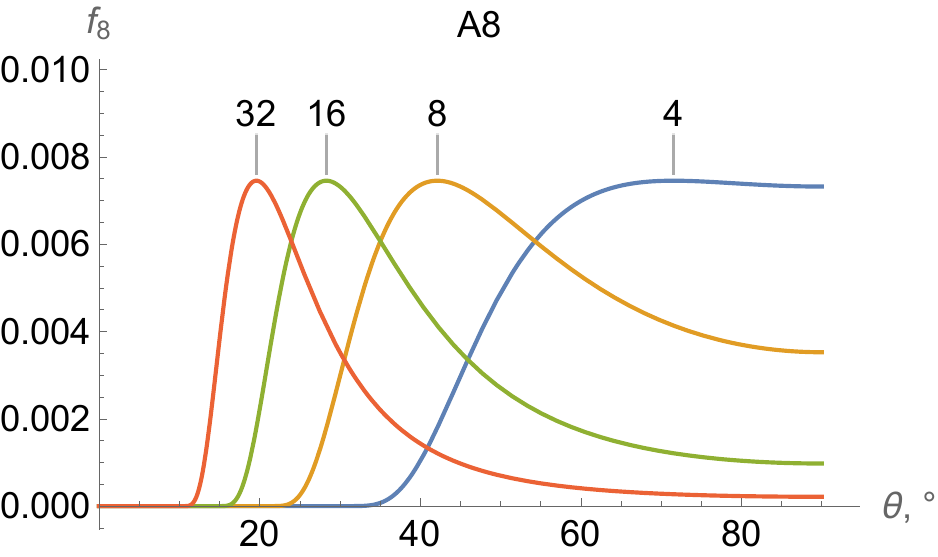}
  \caption{
    Angular distribution of the fast ions in the A1, A2, A3, and A8 models for a set of parameters $R \in {4,8,16,32}$.
  }\label{fig:AngleDistribution_n1238_vs_theta}
\end{figure*}

Examples of the angular distribution of charged particles are shown in Fig.~\ref{fig:AngleDistribution_n1238_vs_theta} for the A1, A2, A3 and A8 pressure models. Comparing these figures, it can be concluded that the width of the angular distribution increases, and the degree of anisotropy in its intuitive interpretation decreases as the index $n$ gets larger.

Knowing the function $F_{n}(X)$, one can find the plasma density distribution:
    \begin{gather}
    \label{003.4:16}
    N =  n_0 f_k(\psi)
    \frac{b}{2n}\left(1-b\right)^{n-1}
    \left(1+(2n-1)\,b\right)
    ,
    \\
    p_{0}/n_{0}
    =
    \int_{0}^{\infty }\varepsilon^{3/2}g(\varepsilon )\,\dif\varepsilon
    \bigm/
    \int_{0}^{\infty }\varepsilon^{1/2}g(\varepsilon )\,\dif\varepsilon
    .
    \end{gather}

In a similar way, it is possible to restore the angular part of the distribution function in model A1:
    \begin{gather}
    F_{1}(X)
    =
    \frac{
        X^2+3 \sqrt{X-1}\, X^2 
        \tan ^{-1}
        \left(\sqrt{X-1}\right)
        -X-2
    }{2\pi  \sqrt{X-1}\, X^2}
    \,
    ,
    \end{gather}
and then the dependence of density on the magnetic field:
    \begin{gather}
    N =  n_0 f_k(\psi)\,
    \frac{1}{2}\, (1-b) (b+3)
    .
    \end{gather}

\section{MHD stabilization by lateral wall}\label{s04}

It is advisable to divide the study of MHD stabilization of plasma in an open trap using a perfectly conducting lateral wall and end MHD stabilizers into three parts. To begin with, this section presents calculations for the case when only the lateral wall is used. The next section \ref{s05} details stabilization by the MHD end stabilizers. Finally, in section \ref{s06} maps of stability zones under the combined action of the lateral wall and end stabilizers are presented.

If the lateral conducting wall is the only means of suppressing MHD instabilities, any  solution of the Sturm-Liouville problem for the LoDestro equation gives the second critical beta, $\beta_{\text{cr}2}$, which determines the lower margin of the upper stability zone $\beta >\beta_{\text{cr}2}$. Results of such solutions are reported below in this section for the magnetic field \eqref{03:11} with mirror ratios selected from the set $M \in \{16,8,4,2\}$ for some most informative combinations of parameters $k\in\{1,2,4,\infty\}$, $q \in \{2,4,8\}$ and $R$. As can be seen in Fig.~\ref{fig:Bv_vs_z_q}, the real magnetic field in the GDT and WHAM devices is better approximated by large values of $M$ and $q$. However, these values are not very convenient for identifying trends caused by changes in the proportion of magnetic mirrors in the total length of the magnetic trap. In addition, these values are reserved for comparison with upcoming calculations with the real magnetic field in the GDT.

Parameter $R$ varies from $R=1.1$ to $R=M$, taking discrete values from some predefined set, usually $R \in \{1.1,1.2,1.5,2,4,8,16\}$, but the maps in the plane $\{R,\beta\}$ are based on a more dense set.

The width of the vacuum gap between the lateral conducting wall and the lateral surface of the plasma column is determined by the ratio of the radius of lateral conducting wall $r_{w}(z)$ to the radius of the plasma column $a(z)$. The ratio $r_{w}(z)/a(z)$ enters the LoDestro equation \eqref{A2:01} through the function
    \begin{gather}
    \label{05:01}
    \Lambda(z) = \frac
    {
        r_{w}^{2}(z) + a^{2}(z)
    }{
        r_{w}^{2}(z) - a^{2}(z)
    }
    .
    \end{gather}
The PEK code accepts as an input parameter the ratio of the wall radius $r_{w0}=r_{w}(0)$ to the radius of the plasma column $a_{0}=a(0)$ in the median plane $z=0$ of an imaginary trap, where the magnetic field is minimal. Calculation was made for a discrete set of $r_{w0}/a_{0}$ values from $\sqrt{501/499}$ to $\infty$, which corresponds to $\lambda_{0}=\Lambda(0)$ values from $500 $ to $1$. 

The results of calculations are presented below for both a proportional lateral wall shape (labelled by Pr) and the straightened one (St). In the first case, the ratio $r_{w}(z)/a(z)=r_{w0}/a_{0}=\const$ is the same in all cross sections $z=\const$ of an imaginary trap, so the function \eqref{05:01} is also constant, $\Lambda(z)\equiv\Lambda_{0}$. The radius of the plasma column in all cases decreases monotonically from the maximum value $a_{0}$ in the median plane $z=0$ through $a_{R}=\sqrt{2}$ at the turning point to the minimum value in the magnetic mirrors $a_{\min}=a(\pm1)=\sqrt{2R/M}$. Note that $a_{0}$ in the theoretical treatment  depends on parameter $R$, radial index $k$ and the value of $p_{0}$ to be given as input parameter at the code start. On the contrary,  in an experimental environment both $r_{w0}$ and $a_{0}$ can be considered as given quantities. If $a_{0}$ is fixed by an external limiter placed in the midplane, then $a_{R}$ and $a_{\min}$ turn out to depend not only on $R$ and $M$, but also on $p_{0}$ (i.e. on $\beta $).

The width of the vacuum gap in the proportional conducting shell decreases monotonically. On the contrary, in a straightened chamber, the radius of the conducting wall is fixed, $r_{z}=r_{w}(0)$, so the vacuum gap increases monotonically as an observation point moves from the center of the trap to the mirror plugs. Thus, for the same ratio $r_{w0}/a_{0}$ average width of the vacuum gap is larger in case of straightened lateral wall. It is therefore the stable zones are always smaller for the straight lateral wall case. This is demonstrated numerically \citep{ZengKotelnikov2024PPCF_66_075020}.

\subsection{Minimal vacuum gap}\label{s04.1}

\begin{figure*}
\centering
\includegraphics[width=\linewidth]{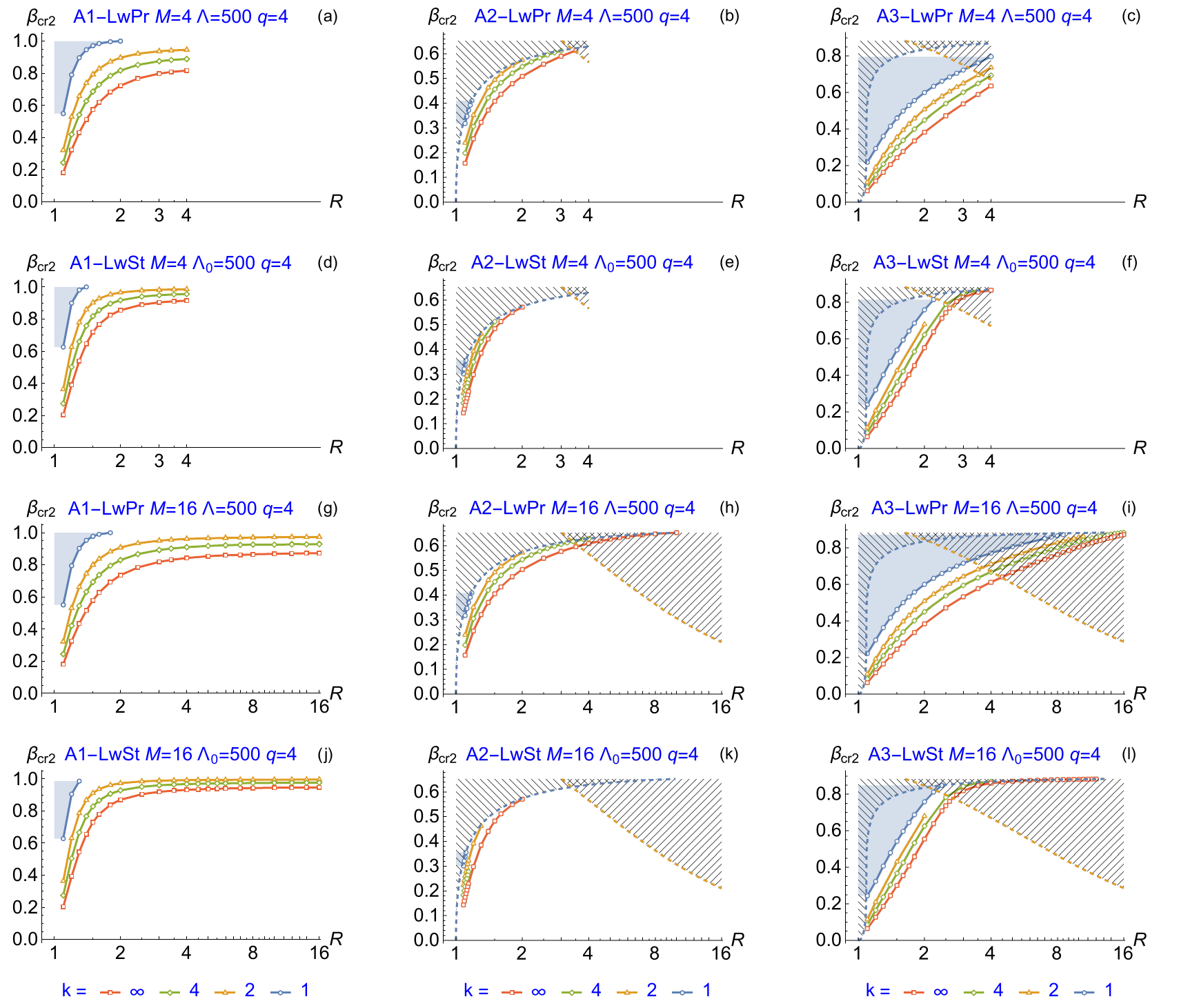}
\caption{
    Second critical beta versus $R$ in the limit $\Lambda\to\infty$ for three pressure models: A1 (left column), A2 (middle), and A3 (right column). 
    Stability zone of ballooning rigid mode for a radial pressure profile with a given index $k$ is located above the margin curve, colored as indicated in the legend under the bottom row. 
}
\label{fig:24-GM2-maxL_A1A2A3_M4M16_q4-LWPrLwSt-beta_vs_R}
\end{figure*}
\begin{figure*}
\centering
\includegraphics[width=\linewidth]{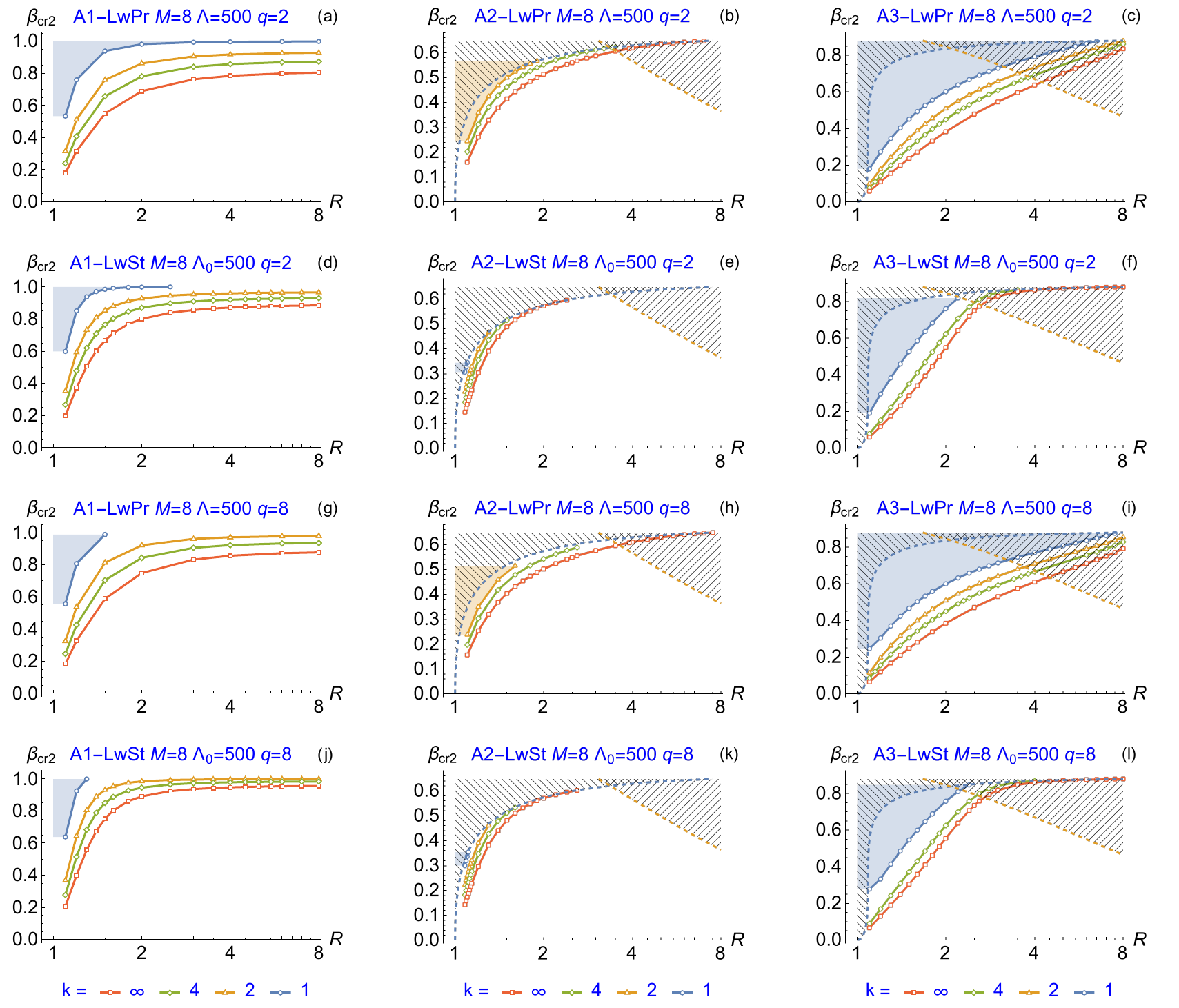}
\caption{
    Second critical beta versus $R$ in the limit $\Lambda\to\infty$ for three pressure models: A1 (left column), A2 (middle), and A3 (right column). 
    Stability zone of ballooning rigid mode for a radial pressure profile with a given index $k$ is located above the margin curve, colored as indicated in the legend under the bottom row. 
}
\label{fig:24-GM2-maxL_A1A2A3_Ms_q2q8-LWPrLwSt-beta_vs_R}
\end{figure*}

There is hardly any need to prove that the stabilizing effect of the lateral conducting wall is maximum when it is located closest to the surface of the plasma column. Therefore, as a first study of a new configuration, it is reasonable to carry out calculations of the critical beta with the conducting wall located as close as possible to the surface of the plasma column. For the pressure model A1, corresponding to the transverse NBI, such calculations are given in Refs.~\citep{Kotelnikov+2023NF_63_066027, ZengKotelnikov2024PPCF_66_075020}. Figures 
\ref{fig:24-GM2-maxL_A1A2A3_M4M16_q4-LWPrLwSt-beta_vs_R}
 and
\ref{fig:24-GM2-maxL_A1A2A3_Ms_q2q8-LWPrLwSt-beta_vs_R}
below are compiled to compare the stabilizing effect of the lateral wall for the three pressure models: A1, A2, A3.
They show a series of graphs for the case $\Lambda_{0}=500$ and illustrate the dependence of $\beta_{\text{cr}2}$ on $R$ and lateral wall shape. Graphs in odd and even rows are drawn for proportional and straightened conducting walls, respectively. 

Both figures confirm the previously discovered strong dependence of the critical beta $\beta_{\text{cr}2}$ on the shape of the radial pressure profile, represented by the index $k$. The region of stability with respect to flute and ballooning disturbances in these figures is located above the marginal beta curve $\beta _{\text{cr}2}(R)$, the color of which is associated with the index $k$ according to the legend under the bottom row of graphs in each figure. The areas of mirror and fire hose instabilities are hatched by the scheme presented in Fig.~\ref{fig:StableZone-A2A3-beta_vs_R}. Minimal area of stability in respect to ballooning modes is shaded with the color of the most upper curve. Most often, this is blue color corresponding to the most smooth radial profile $k=1$, but in graphs~\ref{fig:24-GM2-maxL_A1A2A3_Ms_q2q8-LWPrLwSt-beta_vs_R}(b,h) this blue zone is invisible since corresponding blue curve is too short because it crosses lower margin of the mirror mode. The PEK code quits above the mirror threshold where paraxial equilibrium is not possible.

Previously, it was found for model A1 \citep{Kotelnikov+2023NF_63_066027, ZengKotelnikov2024PPCF_66_075020} that when the end MHD stabilizers are switched off, the critical value $\beta_{\text{cr}2}$ depends only very weakly on both the mirror ratio $ M$ and index $q$, which controls the axial profile of the magnetic field. For this reason, in Fig.~\ref{fig:24-GM2-maxL_A1A2A3_M4M16_q4-LWPrLwSt-beta_vs_R}, illustrating the dependence on the parameter $M$, only graphs with index $q=4$ are kept, and the set of values $M$ is reduced to two, $M=4$ and $M=16$. Similarly, in Fig.~\ref{fig:24-GM2-maxL_A1A2A3_Ms_q2q8-LWPrLwSt-beta_vs_R}, illustrating the dependence on $q$, only graphs with the mirror ratio $M=8$ and a pair of index values $q$ are kept, namely: $q=2$ and $q=8$. In general, all these graphs only confirm the conclusion about the weak influence of the magnetic field on the stabilizing properties of the lateral conducting wall within the same model of anisotropic pressure.

However, the differences between the three studied anisotropic pressure models A1, A2 and A3 are visible to the naked eye. These conclusions partially coincide with those of Kesner \citep{Kesner1985NF_25_275}. He stated that the critical beta was insensitive to the axial profile of the magnetic field, but did not specify the value of the index $n$ in Eq.~\eqref{003:02} used in his calculations.

In model A2, the stabilization boundary of ballooning instability is closely adjacent to the threshold of mirror instability, so that the region of joint stability $\beta_{\text{cr}2} < \beta < \beta_{\text{mm}} $ turns out to be very narrow. In model A3 this region is noticeably wider, especially in the region $R\lesssim2$. In both models, the zone of stabilization of ballooning modes partially overlaps the region of fire hose instability at $R\gtrsim4$.

Thus, it can be concluded that the stability properties of an anisotropic plasma are very sensitive to details of the axial pressure profile.

In the most general terms, the last statement can hardly be disputed. At the same time, it should be noted that model A2 apparently does not adequately describe a real experiment. As follows from the discussion of Eq.~\eqref{003.3:01}, the point of origin of mirror instability in this model coincides with the turning point $b=1$ of fast ions, since the derivative $\tparder{p_{\bot}}{b}$ in this point experiences a jump as a result of mathematical idealization. A quick look at Fig.~\ref{fig:AngleDistribution_n1238_vs_theta} shows that both the increase in anisotropy and the decrease in NBI tilt angle make it more difficult to stabilize an anisotropic plasma.

Comparison of graphs in odd and even rows of Figs.~\ref{fig:24-GM2-maxL_A1A2A3_M4M16_q4-LWPrLwSt-beta_vs_R} 
and~\ref{fig:24-GM2-maxL_A1A2A3_Ms_q2q8-LWPrLwSt-beta_vs_R} 
reveals that the shape of the conducting chamber significantly deforms shape of the $\beta_{\text{cr}2}(R)$ curves in case of oblique NBI simulated by the models A2 and A3. For the model A1, Ref.~\citep{ZengKotelnikov2024PPCF_66_075020} made the opposite conclusion that the effect of the shape of the conducting chamber is insignificant. The apparent contradiction can be explained by the fact that the width of the gap between the conducting wall and the plasma at the location of  the pressure peak plays a decisive role. In the model A1 (transverse NBI), the pressure peak is located in the midplane of the trap. For the same value of parameter $\Lambda_{0}$, the vacuum gap in the midplane  is the same for the proportional and straightened chambers, so the difference in the gap width in other sections of the trap is not very significant. In the case of A2 and A3 models (oblique NBI), the peak is located outside the midplane. The size of the gap at the location of the peak in the strengthened chamber will be larger than in the proportional chamber; accordingly, the stabilizing effect of the conducting chamber will be less.

Comparison of graphs in successive columns of Figs.~\ref{fig:24-GM2-maxL_A1A2A3_M4M16_q4-LWPrLwSt-beta_vs_R} 
and 
\ref{fig:24-GM2-maxL_A1A2A3_Ms_q2q8-LWPrLwSt-beta_vs_R} 
shows that the critical beta $\beta_{\text{cr}2}$ is definitely smaller in the case of oblique injection (which is good), but the stability zone disappears at large values of parameter $R$, since the plot of $\beta_{\text{cr}2}(R)$ crosses both the mirror instability threshold $\beta_{\text{mm}}(R)$ and the fire hose instability threshold $\beta_{\text{fh}}(R)$ at about $R\approx 3.5\text{–}4$. Considering that Eq.~\eqref{003.4:15} relates parameter $R$ to the injection angle $\theta_{\text{inj}}$, one obtains $\theta_{\text{inj}}\approx 40^{\circ}$ in model A2 and $\theta_{\text{inj}}\approx 47^{\circ}$ in model A3. In a straightened chamber, the $\beta_{\text{cr}2}$ curve intersects the $\beta_{\text{mm}}$ curve at a lower value of the parameter $R\approx 2$, which corresponds to $ \theta_{\text{inj }}\approx 56^{\circ}$ in model A2 and $\theta_{\text{inj}}\approx 70^{\circ}$ in model A3.

The second observation is that the straightened chamber stabilizes the smoothest radial pressure profile with index $k=1$ noticeably worse, in the sense that the range of values of the $R$ parameter at which this profile can be stabilized is noticeably narrower compared to the proportional chamber. In Section \ref{s06}, it is shown that in the presence of the end MHD stabilizers, on the contrary, the smoothest radial profile is the easiest to stabilize because the unstable zones on the stability maps in Figs.~\ref{fig:24-GM2-minL_A1A2A3_M16_q2q4q8-CwPr-beta_vs_R_graphs9}–\ref{fig:GM2-A3-CWSt_beta_vs_rw_setM16q4} are the smallest for the $k=1$ radial profile.

\begin{figure*}
  \centering
  \includegraphics[width=\linewidth]{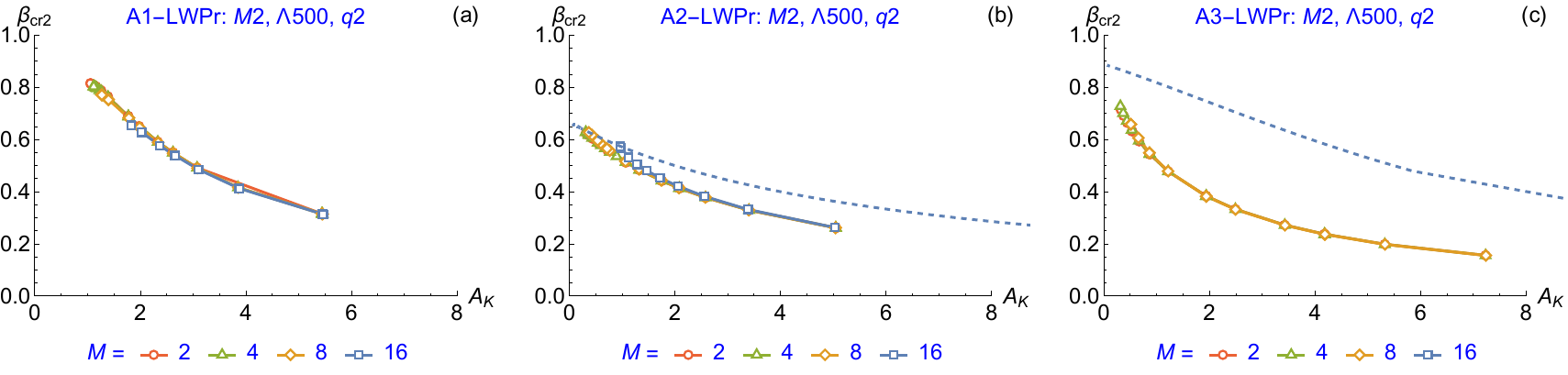}
  \caption{
    Second critical beta $\beta_{\text{cr}2}$ versus Kesner's degree of anisotropy $A_{\text{K}}$ defined by Eq.~\eqref{003.3:02} as in Ref.~\citep{Kesner1985NF_25_275} for three models of anisotropic pressure and four mirror ratios, $M\in\{4,8,16\}$. Dashed curve shows the mirror-mode thresholds. Compare with Fig.~3\textsuperscript{\citep{Kesner1985NF_25_275}} in Ref.~{\citep{Kesner1985NF_25_275}}.
  }
  \label{fig:23-GM2-A1A2A3-LWPr_beta_vs_A_graphs3}
\end{figure*}
It is useful to compare the above depicted calculations with the results of predecessors, in particular, with the works of Kesner and his co-authors \citep{
     Kesner1985NF_25_275,
     LiKesnerLane1985NF_25_907,
     LiKesnerLane1987NF_27_101,
     LiKesnerLoDestro1987NF_27_1259
}. In all these papers, the authors analyze the plasma model with sharp boundary, which corresponds to the infinite index $k=\infty$. Direct comparison is possible only with the first of the enlisted publications, and then only with certain reservations. In other works, the differences in the formulation of the problem are too great.

In his first work \citep{Kesner1985NF_25_275}, Kesner approximated the vacuum magnetic field with a parabola. In the model of magnetic field \eqref{03:11}, a parabola can approximate the field near the midplane in case of $q=2$. Therefore, results can be compared in cases where the pressure peak is concentrated near the median plane, that is, at small values of parameter $R\in \{1.1,1.2,1.3\}$. The necessary data can be extracted from\footnote{
    A reference to a figure in another article is hereinafter supplemented with a superscript indicating the reference number in the bibliography.
} Fig.~2\textsuperscript{\citep{Kesner1985NF_25_275}}, which shows the calculation results (in our terms) for the model A1-LwPr and zero vacuum gap. Comparing these data with figures 2\textsuperscript{\citep{Kotelnikov+2022NF_62_096025}}(a,d,g,j), one can see quite satisfactory agreement.

It is more difficult to compare the results of calculations in the case of oblique injection. Fig.~3\textsuperscript{\citep{Kesner1985NF_25_275}} shows the graph of $\beta_{\text{cr}2}$ depending on the Kesner's anisotropy degree \eqref{003.3:02} in the range from $A_{\text{K}}=1$ to $A_{\text{K}} =6$. The explanation to that figure indicates that the plasma occupies the entire length of the trap up to the magnetic mirrors, i.e.\ $R=M$, but the magnetic field was still approximated by a parabola. The values of the parameters $M$, $R$, $n$ are not specified in the article, however, taking into account the definition \eqref{003.3:02} of the anisotropy parameter in that article, one can assume that there should be $n\gg M$ in order to have such values of $A_{\text{K}}$. Formally, this means that comparison with my calculations for $n=2$ (i.e., model A2) and $n=3$ (model A3) is impossible.

Alternative possibility to ``engineer'' a large parameter $A_{\text{K}}$ involves the simultaneous limit $R\to1$ and $\beta \to 0$. Parameter $A_{\text{K}}$ may well fall into the range $1\text{–}6$ for $n=2\text{–}3$ if $\beta\ll1$ and $R=1.1\text{–}2$. In order to draw Fig.~\ref{fig:23-GM2-A1A2A3-LWPr_beta_vs_A_graphs3} in the $\{A_{\text{K}}, \beta\}$ coordinates, suitable for comparison with Fig.~3\textsuperscript{\citep{Kesner1985NF_25_275}} in Ref.~\citep{Kesner1985NF_25_275}, $\beta_{\text{cr}2}$ was calculated for a set of discrete values of $R$ with four values of the mirror ratio $M\in \{2, 4, 8, 16\}$. Comparing Fig.~\ref{fig:23-GM2-A1A2A3-LWPr_beta_vs_A_graphs3} and Fig.~3\textsuperscript{\citep{Kesner1985NF_25_275}}, one can see some qualitative discrepancies. Most evident of them is that in Fig.~\ref{fig:23-GM2-A1A2A3-LWPr_beta_vs_A_graphs3}(b,c) the width of the stable zone between the $\beta_{\text{cr}2}$ and $\beta_{\text{mm}}$ curves decreases as $A_{\text{K}}$ becomes smaller whereas Fig.~\ref{fig:23-GM2-A1A2A3-LWPr_beta_vs_A_graphs3} demonstrates an opposite tendency.

It does not follow from this fact that Kesner's results are wrong. In fact, I don't quite understand his assumptions because some important details are missed. In addition, the choice of parameters $A_{\text{K}}$ and $\beta$ for Fig.~3\textsuperscript{\citep{Kesner1985NF_25_275}} axes is unsuccessful, since parameter $A_{\text{K}}$ itself depends on $\beta$.

\subsection{Effect of vacuum gap}\label{s04.2}

\begin{figure*}
  \centering
\includegraphics[width=\linewidth]{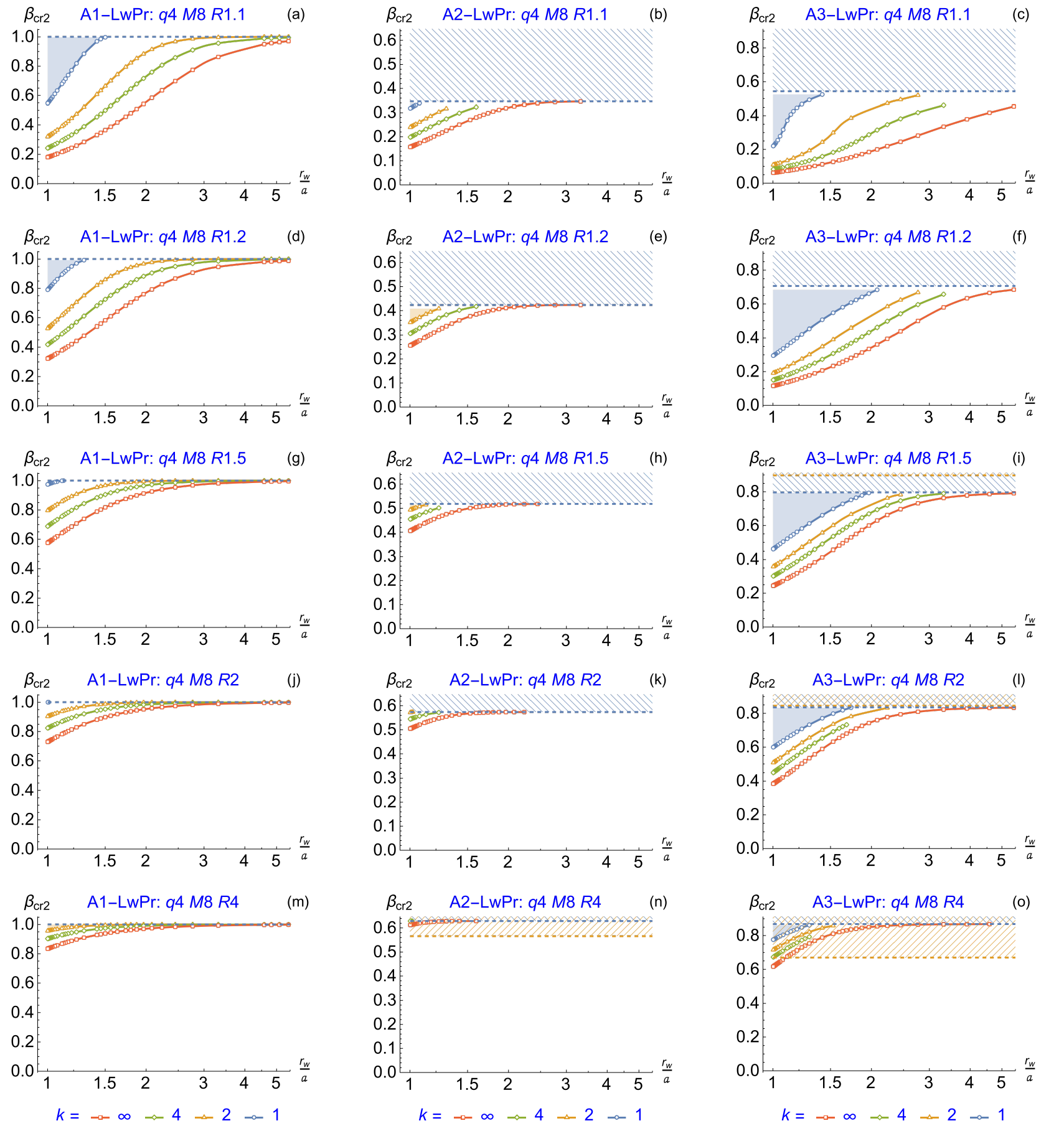}
  \caption{
    Stability maps for the LwPr configuration and three pressure models: A1 (left column), A2 (middle column), and A3 (right column). Second critical beta is drawn as a function of $r_{w}/a$ for the model magnetic field \eqref{03:11} with index $q=4$, set of mirror ratios $R\in \{1.1,1.2,1.5,2,4,8\}$ at the turning point, and fixed mirror ratio $M=8$.
    Stable zone for a radial pressure profile with an index $k$ is located above the curve $\beta_{\text{cr}2}$ colored according to the legend under the bottom row.
    Compare with Fig.~\ref{fig:GM2-A1A2A3_N8_q4-LWSt_beta_vs_rw_graphs15}.
    }
    \label{fig:GM2-A1A2A3_N8_q4-LWPr_beta_vs_rw_graphs15}
\end{figure*}

\begin{figure*}
  \centering
\includegraphics[width=\linewidth]{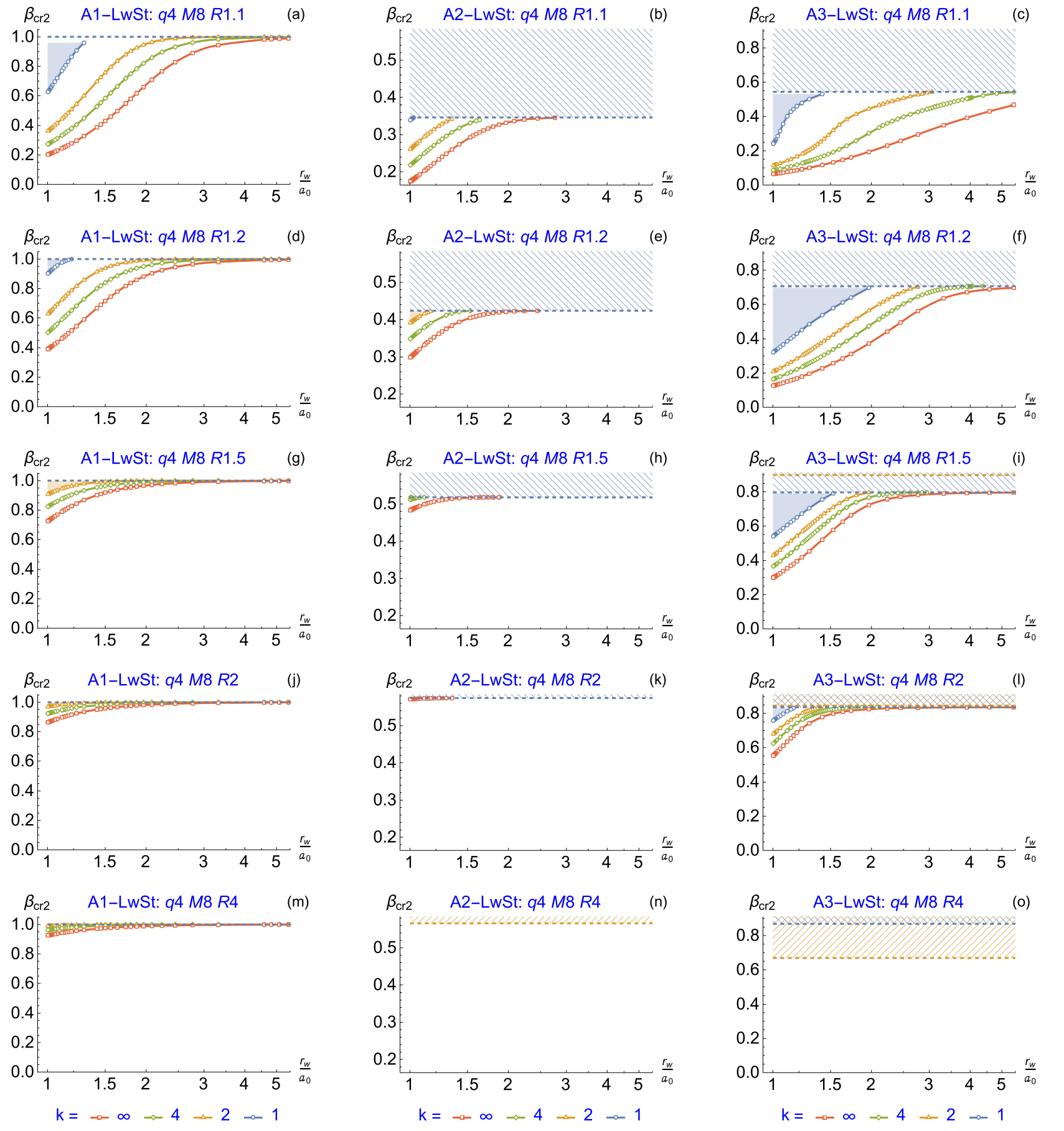}
  \caption{
    Stability maps for the LwSt configuration and three pressure models: A1 (top row), A2 (second row), and A3 (bottom row). Second critical beta is drawn as a function of $r_{w}/a$ for the model magnetic field \eqref{03:11} with index $q=4$, set of mirror ratios $R\in \{1.1,1.2,1.5,2,4,8\}$ at the turning point, and fixed mirror ratio $M=8$.
    Stable zone for a radial pressure profile with an index $k$ is located above the curve $\beta_{\text{cr}2}$ colored according to the legend under the bottom row.
    Compare with Fig.~\ref{fig:GM2-A1A2A3_N8_q4-LWPr_beta_vs_rw_graphs15}.
    }
  \label{fig:GM2-A1A2A3_N8_q4-LWSt_beta_vs_rw_graphs15}
\end{figure*}

The width of the vacuum gap between the lateral surface of the plasma column and the inner surface of the conducting chamber of a selected type is determined by the parameter
\begin{gather*}
r_{w0}/a_{0}=\sqrt{(\Lambda_{0}+1)/(\Lambda_{0}-1)}
. 
\end{gather*}
In previous section \ref{s04.1} it was shown that in the limit $r_{w0}/a_{0}\to1$ the critical beta remains essentially unchanged when both the mirror ratio $M$ and the index $q$ change. This conclusion holds for any value of $r_{w0}/a_{0}$, so it is sufficient to show only for one (say, average) value of the index and one value of the mirror ratio. In addition, the range of changes in the parameter $R$ is limited below by the interval $R\in[1.1\ldots 4]$, since at $R\gtrsim4$ the margin of stability with respect to ballooning oscillations lies inside the fire hose instability zone.

Figures 
\ref{fig:GM2-A1A2A3_N8_q4-LWPr_beta_vs_rw_graphs15} 
and 
\ref{fig:GM2-A1A2A3_N8_q4-LWSt_beta_vs_rw_graphs15}
contain series of graphs of $\beta_{\text{cr}2}$ versus ratio $r_{w0}/a_{0}$ at fixed mirror ratio $M = 8$ and fixed axial profile index $q=4$. Fig.~\ref{fig:GM2-A1A2A3_N8_q4-LWPr_beta_vs_rw_graphs15} is compiled for proportional chamber LwPr, and 
Fig.~\ref{fig:GM2-A1A2A3_N8_q4-LWSt_beta_vs_rw_graphs15} repeats the same graphs for the LwSt configuration.

Comparison of graphs within any row demonstrates a strong dependence of $\beta_{\text{cr}2}$ on pressure model. It can be seen that for $r_{w}/a\approx2$ the lowest critical beta is much smaller in the case of the oblique NBI imitated by the model A2 (middle column) compared to the transverse NBI described by model A1 (left column). In other words, with an oblique NBI it is easier to achieve a stable plasma confinement mode. The effect is even more significant in model A3 (right column). However, it is very difficult to stabilize a plasma with a maximally smooth radial pressure profile $k=1$ during oblique injection. In model A2, the blue curve corresponding to the index $k=1$ is completely absent from most graphs. In addition, in model A2, the boundary of the stability zone $\beta =\beta_{\text{cr}2}$ rests on the threshold of mirror instability $\beta =\beta_{\text{mm}}$ even for a not very large ratio $ r_{w}/a$. As it is noted above, model A2 is not realistic enough. Model A3 predicts a wider range of $r_{w}/a$ values at which stabilization by a properly designed conducting chamber is possible.


\section{MHD stabilization by end anchors}\label{s05}

A proven method for suppressing MHD instabilities in an axially symmetric mirror trap is to attach end MHD anchors to the central mirror cell on the side of each of the two magnetic mirrors \citep{Ryutov+2011PoP_18_092301}. The PEK package simulates MHD end stabilizers using boundary conditions of the form \eqref{A2:11S} in the $z=z_{E}$ plane, where an imaginary conducting end plate is located. The further such a plate is placed from the center of the trap, the less its stabilizing effect \citep{ZengKotelnikov2024PPCF_66_075020}.

If $z_{E}>1$, the end plate is located behind the magnetic mirror and may not be virtual, but real, as in some experiments at the GDT installation \citep{Soldatkina+2017PoP_24_022505, Soldatkina+2020NF_50_8}. It is interesting and to some extent unexpected that such a plate, contrary to well-founded fears, did not lead to degradation of plasma parameters, even if it was placed relatively close to the neck of the magnetic plug. Variant $z_{E}>1$ simulates a terminal MHD stabilizer of the cusp type \citep{Taylor1963PhysFluids_6_1529, Logan1980CPPCF_5_271, Logan1981CPPCF_6_199, Baker+1984PhysFluids_27_22723, Anikeev+1887PoP_4_347, LI+2023PST_25_025102}.

If $z_{E}<1$, the imaginary end plate is located in front of the magnetic plug. In the problem of stabilizing the rigid ballooning mode, we can assume that its role is played by a limiter in the form of a ring, which is used in the vortex confinement method \citep{Bagryansky+2003FST_43_152, Bagryansky+2007FST_51_340, Bagryansky+2011FST_59_31, Beklemishev2008AIP_1069_3}.

Calculation of the stability margin that is created by one or another design of the end MHD anchor is a separate task. It is beyond the scope of this article (see, for example, \citep{NagornyiStupakov1984SovJplasPhys_10_275, KuzminLysyanskij1990FizPlaz_16_1001}). The end plate, located behind the magnetic plug, imitates an MHD anchor, which has a smaller stability margin. On the contrary, the end plate installed in front of the magnetic plug imitates an MHD anchor, which has a larger stability margin.

According to the historically established classification inside the PEK package, the variants $z_{E}=1$, $z_{E}>1$ and $z_{E}<1$ are designated as Cw (Combined wall), Bw (Blind wall) and Rw (Ring wall), respectively, although in the current version of the PEK package, the calculation of all three options is carried out by a common module. To limit the number of figures, only the case $z_{E}=1$ (Cw) is presented below, when an imaginary conducting plate is located in the neck of the magnetic plug.

The absence of a lateral conducting shell corresponds to the limit $r_{w}/a_{0}\to\infty$ (that is, $\Lambda\to1$). In this limit, the shape of the conducting shell obviously does not matter and, in this sense, the options CwPr, CwSt (as well as BwPr, BwSt, RwPr, RwSt) are equivalent. The PEK package in this limit finds at most one root, but now, in contrast to section \ref{s04}, this is the upper margin $\beta_{\text{cr}1}$ of the lower stability zone.

\begin{figure*}
\centering
\includegraphics[width=\linewidth]{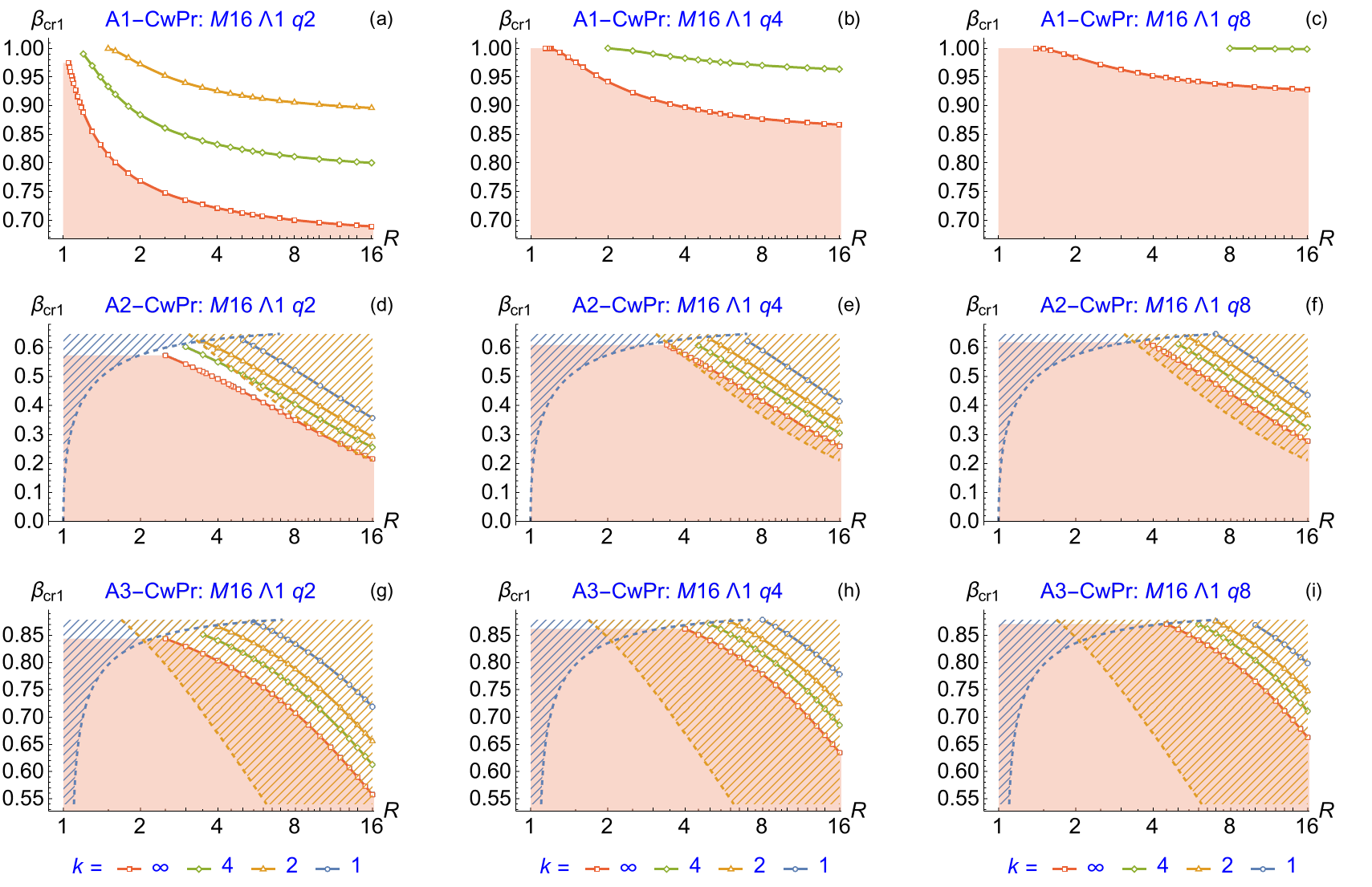}
\caption{
    First critical beta $\beta_{\text{cr}1}$ for the CwPr configuration and three pressure models: A1 (upper row), A2 (middle row), and A3 (bottom row) at mirror ratio $M=16$ versus parameter $R$ in the limit $\Lambda\to1$. 
    The stability zone of rigid ballooning perturbation for a radial pressure profile with a given index $k$ is located below the curve, colored as indicated in the legend under the bottom row of graphs. 
    }\label{fig:24-GM2-minL_A1A2A3_M16_q2q4q8-CwPr-beta_vs_R_graphs9}
\end{figure*}
\begin{figure*}
\centering
\includegraphics[width=\linewidth]{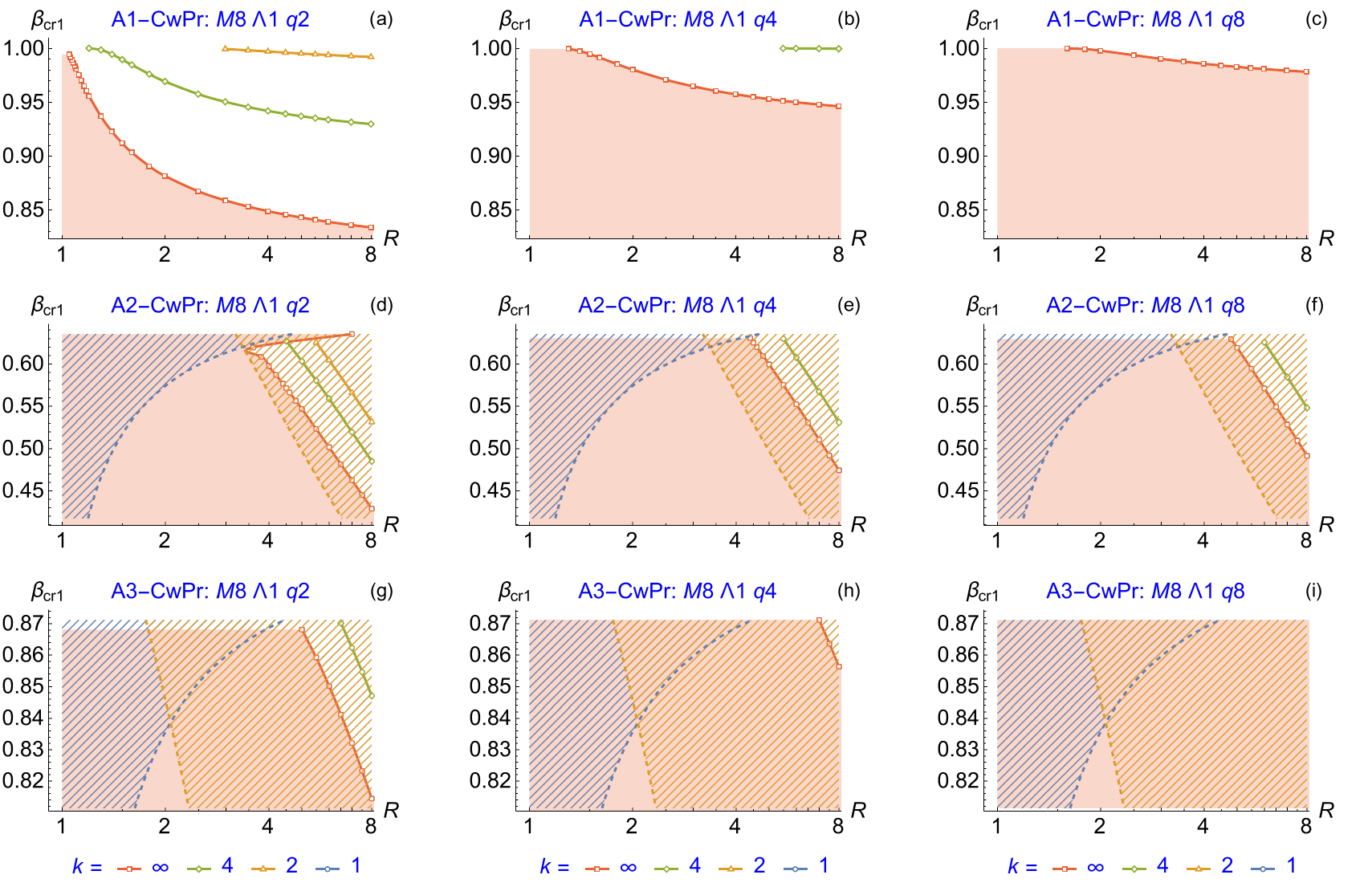}
\caption{
    Same maps as in Fig.~\ref{fig:24-GM2-minL_A1A2A3_M16_q2q4q8-CwPr-beta_vs_R_graphs9} but for mirror ratio $M=8$. 
}
\label{fig:24-GM2-minL_A1A2A3_M8_q2q4q8-CwPr-beta_vs_R_graphs9}
\end{figure*}

\begin{figure*}
\centering
\includegraphics[width=\linewidth]{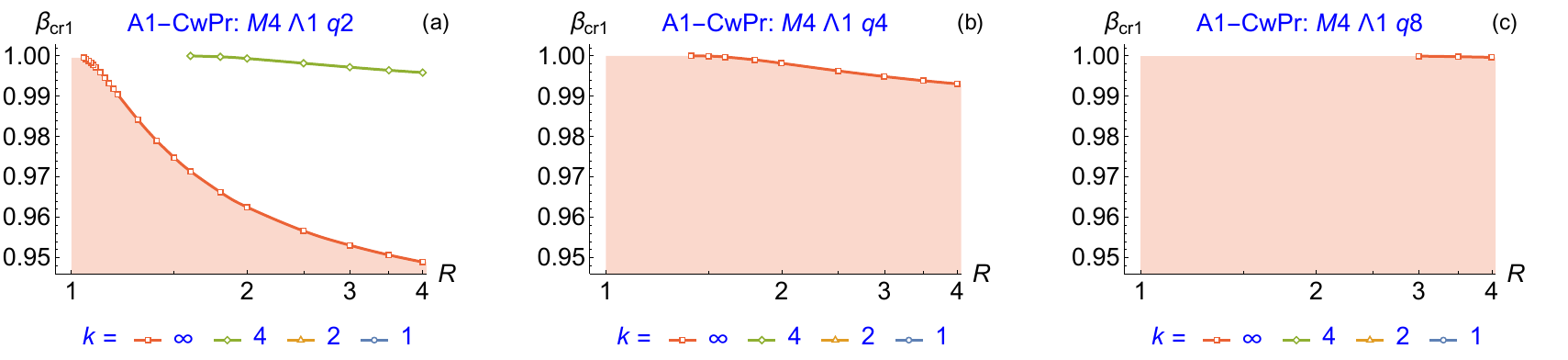}
\caption{
    Same maps as in Fig.~\ref{fig:24-GM2-minL_A1A2A3_M16_q2q4q8-CwPr-beta_vs_R_graphs9} but for mirror ratio $M=4$. The rows A2-CwPr and A3-CwPr are not shown as they don't have unstable zone for rigid ballooning modes.
}
\label{fig:24-GM2-minL_A1A2A3_M4_q2q4q8-CwPr-beta_vs_R_graphs9}
\end{figure*}

Series of figures \ref{fig:24-GM2-minL_A1A2A3_M16_q2q4q8-CwPr-beta_vs_R_graphs9}, \ref{fig:24-GM2-minL_A1A2A3_M8_q2q4q8-CwPr-beta_vs_R_graphs9}, and~\ref{fig:24-GM2-minL_A1A2A3_M4_q2q4q8-CwPr-beta_vs_R_graphs9} illustrates the dependence of $\beta_ {\text{cr1}}$ on parameter $R$ for three models of anisotropic pressure A1, A2, A3, three values of the mirror ratio $M\in\{16,8,4\}$ and three magnetic field profiles, corresponding to index values $q\in\{2,4,8\}$. Analysis of the graphs in these figures shows that when using MHD stabilizers, all previously made conclusions for the case of lateral wall stabilization have to be reversed.

The stability zone for the radial profile with index $k$ is located below the curve $\beta=\beta_{\text{cr1}}(R)$ colored as indicated in the legend under the bottom row in each figure. Absence of a curve for a certain range of $R$ means that the corresponding radial profile $k$ is stable over the entire interval of betas allowed by the inequalities $0<\beta < 1$ (in case of transverse NBI as described by the A1 model) or $0<\beta<\beta_{\text{mm}}$ below the threshold of mirror instability (in case of oblique NBI simulated by the models A2, A3, \ldots). The areas, where the mirror and fire hose modes are unstable, are shaded in blue and brown color, respectively. Computation was performed for a wider set of mirror ratios $M$ then mentioned above but no $\beta_{\text{cr}1}$ was found for $M\lesssim 4$ .

Comparison of \ref{fig:24-GM2-minL_A1A2A3_M16_q2q4q8-CwPr-beta_vs_R_graphs9}, \ref{fig:24-GM2-minL_A1A2A3_M8_q2q4q8-CwPr-beta_vs_R_graphs9}, and~\ref{fig:24-GM2-minL_A1A2A3_M4_q2q4q8-CwPr-beta_vs_R_graphs9}  reveals a strong dependence of the ballooning instability threshold on mirror ratio $M$. This drastically differs the lower zone of stability from the upper one which  exists exclusively due to lateral conducting wall as discussed in Section \ref{s04}. The lower stability zone tends to expand as the mirror ratio becomes smaller. This fact is quite consistent with the intuitive expectations that a decrease in $M$ improves the contact of the end MHD stabilizer with the central section of the open trap.


These figures show that the anisotropic pressure models have qualitative and quantitative differences. It is especially noticeable when passing from transverse injection (model A1) to oblique injection (models A2 and A3). The difference between models A2 and A3 is limited in quantitative indicators, although it is quite noticeable. It can also be noted that the influence of the radial profile on the value of $\beta _{\text{cr1}}$ in model A1 is more noticeable than in models A2 and A3. This statement follows from the fact that the distance between curves of different colors in the graphs in the first row of each of the figures is \ref{fig:24-GM2-minL_A1A2A3_M16_q2q4q8-CwPr-beta_vs_R_graphs9}, \ref{fig:24-GM2-minL_A1A2A3_M8_q2q4q8-CwPr-beta_vs_R_graphs9}, and~\ref{fig:24-GM2-minL_A1A2A3_M4_q2q4q8-CwPr-beta_vs_R_graphs9} is significantly larger than in the second and third rows. The absence of a blue, yellow or green curves means that the corresponding fairly smooth radial profiles are stable with respect to ballooning perturbations of plasma equilibrium. Again, this effect is more pronounced in the case of transverse NBI, whereas in the case of oblique injection, the fire hose instability threshold $\beta_{\text{fh}}$ in terms of beta value is often lower than the ballooning instability threshold $\beta_{\text{cr1} }$.


Comparison of graphs within every row of any of the three figures shows that the shape of the axial profile of the magnetic field has a significant effect on the value of $\beta_{\text{cr}1}$. This fact is most noticeable in the case of transverse injection (model A1, first row of each figure) and also significantly distinguishes stabilization using end MHD stabilizers from stabilization using a lateral wall. The general tendency is that the narrowing and steepening of magnetic mirrors expands the lower stability zone.

    
With transverse injection, the entire range $0<\beta<1$ can be made stable for any radial pressure profile if $M$, $R$ are sufficiently low. With oblique injection, the $\beta =1$ limit on transverse equilibrium is unattainable due to the development of mirror and fire hose instabilities. However, the last statement regarding the mirror instability was proven only within the paraxial approximation. A number of publications give reason to assume that the threshold of shear instability can be exceeded in nonparaxial open traps \citep{Lansky1993BINP_96, Lotov1996PoP_3_1472, Kotelnikov+2010PhysRevE_81_067402, Kotelnikov2011FST_59_47, Beklemishev2016PoP_23_082506}. As to fire hose instability, to the best of my knowledge, it was never identified in the mirror traps.

Comparing Figs.~\ref{fig:24-GM2-maxL_A1A2A3_M4M16_q4-LWPrLwSt-beta_vs_R},
\ref{fig:24-GM2-maxL_A1A2A3_Ms_q2q8-LWPrLwSt-beta_vs_R}
from section \ref{s04.1} with Fig.~\ref{fig:24-GM2-minL_A1A2A3_M16_q2q4q8-CwPr-beta_vs_R_graphs9}, \ref{fig:24-GM2-minL_A1A2A3_M8_q2q4q8-CwPr-beta_vs_R_graphs9}, and~\ref{fig:24-GM2-minL_A1A2A3_M4_q2q4q8-CwPr-beta_vs_R_graphs9} 
it is easy to notice that as $R$ increases, the boundaries of the lower and upper stability zones shift in opposite directions, and in such a way that both zones contract. Note also that in all the figures the blue curves are located above the yellow ones, the yellow ones are above the green ones, and the green ones are above the red ones. In other words, the boundary of both the upper stability zone and the boundaries of both the upper stability zone and that of the lower stability zone shift downward as the radial pressure profile steepens.

But if for the upper zone this order means that it becomes wider as the radial profile steepens, then the expansion of the lower zone, on the contrary, occurs as the radial profile is smoothed.
\section{Combined MHD stabilization}\label{s06}

Taking into account the existence of upper and lower stability zones when the two stabilization methods are applied separately, which were successively described in sections \ref{s04} and \ref{s05}, it is easy to believe that with the simultaneous application of these two stabilization methods, both stability zones will be preserved. The first indication of the presence of two zones of stability can be found in the works of D'Ippolito and Hafizi \citep{DIppolitoHafizi1981PF_24_2274} and D'Ippolito and Myra \citep{DIppolitoMyra1984PF_27_2256}. After the author's works \citep{Kotelnikov+2021PST_24_015102, Kotelnikov+2022NF_62_096025, Kotelnikov+2023NF_63_066027, ZengKotelnikov2024PPCF_66_075020} the simultaneous existence of the two stability zones has become a proven fact based on the examples of isotropic plasma and anisotropic plasma in the special case of transverse NBI. The lower zone exists at low plasma pressure, at $0<\beta<\beta_{\text{cr}1}$, and the second - at high pressure, at $\beta_{\text{cr}2}<\beta <$1 . These two zones merge for larger $\Lambda$, providing overall stability for any beta in the range $0<\beta<1$. It will be shown below that the last statement should be corrected in the case of inclined NBI.


When the lateral wall and end stabilizers act together, PEK often does not find solutions for the set of parameters discussed in the previous sections. This could mean either that the RBM perturbations are stable over the entire range of available beta values below the mirror instability threshold, or that there is an error in the program. Therefore, calculations in the Cw, Bw and Rw modes usually began with the case $R=M$, when the instability zone has maximum dimensions for a fixed set of parameter values $k$, $q$, $M$, $z_{E}$. The remaining part of the article also begins with the case $R=M$ in Cw configuration. The results of calculations in the Bw and Rw configurations are omitted below because conclusion from their analysis is quite banal. The Rw configuration is more stable than the Cw configuration, and the Bw configuration, on the contrary, is less stable in the sense that the instability zone is larger than in the Cw configuration.

\subsection{Wide pressure peak, $M=R$}
\label{s6.1}


\begin{figure*}
\includegraphics[width=\linewidth]{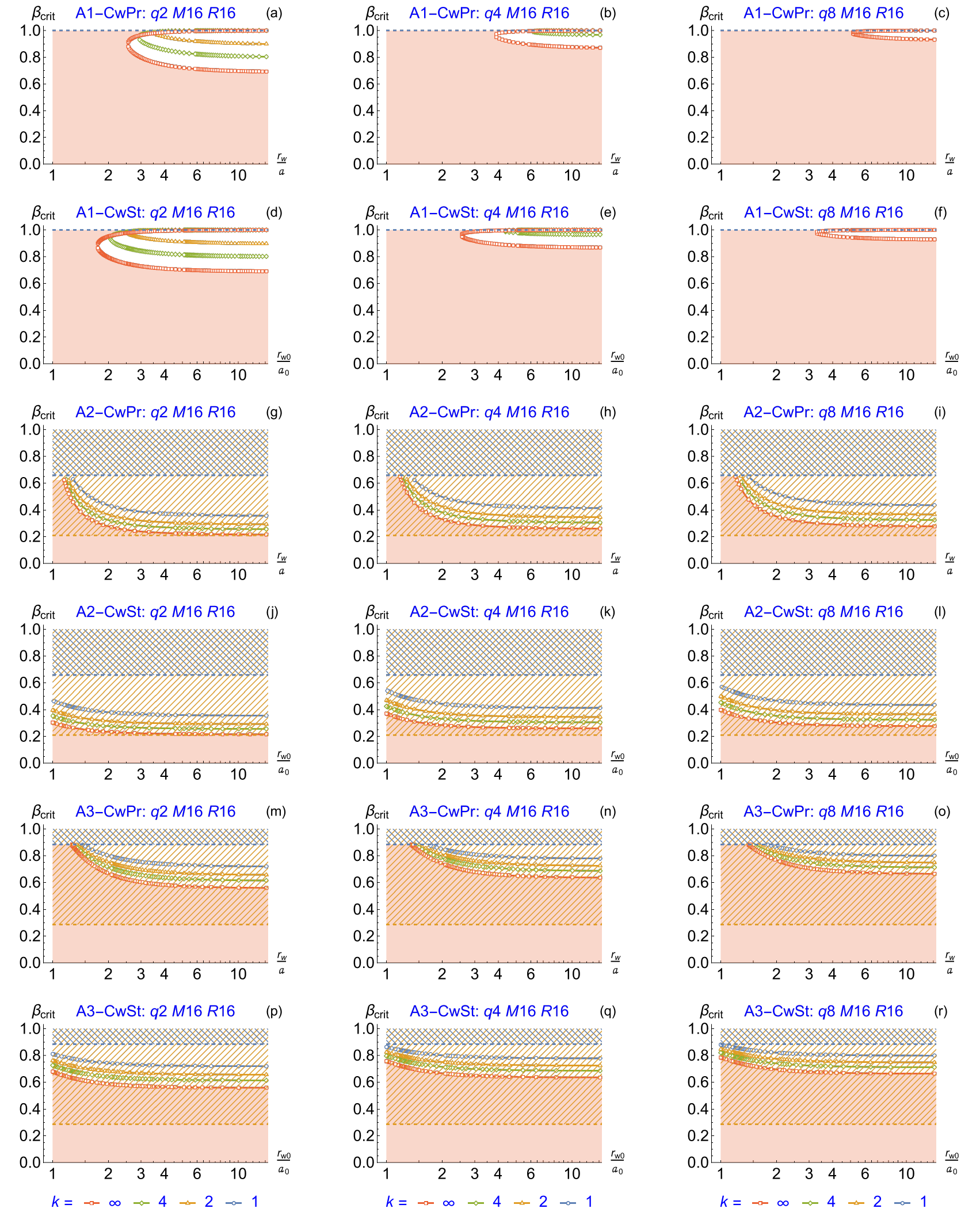}
\caption{
    Stability maps for model magnetic field \eqref{03:11} and anisotropic plasma pressure models \eqref{003.1:01} (1st  and 2nd rows), \eqref{003.2:01} (3rd and 4th rows), and \eqref{003.3:01} (5th and 6th rows) at combined MHD stabilization by lateral wall and end MHD anchors, $q\in\{2, 4, 8\}$, $M=R=16$. 
    The unstable RBM zone is located between the lower $\beta_{\text{cr}1}(r_{w}/a_{0})$ and upper branches $\beta_{\text{cr}2}(r_{w}/a_{0})$ of every curve. 
    Correspondence of the index $k$ to the color of margin curves is shown at the bottom of the figure.
    Shaded common zone of RBM stability lays to the left of the margin curve for the most steep radial pressure profile ($k = \infty $).
    The unstable zones for the fire hose modes and mirror modes shaded in orange and blue, respectively.
    The left column shows maps for a ``parabolic'' magnetic field with the index $q=2$, and the right column shows the maps for the ``quasi-flat hole'' magnetic field with $q=8$.
    The odd and even rows contain maps for the CwPr and CwSt configurations, respectively.
}
\label{fig:M16R16}
\end{figure*}
\begin{figure*}
\includegraphics[width=\linewidth]{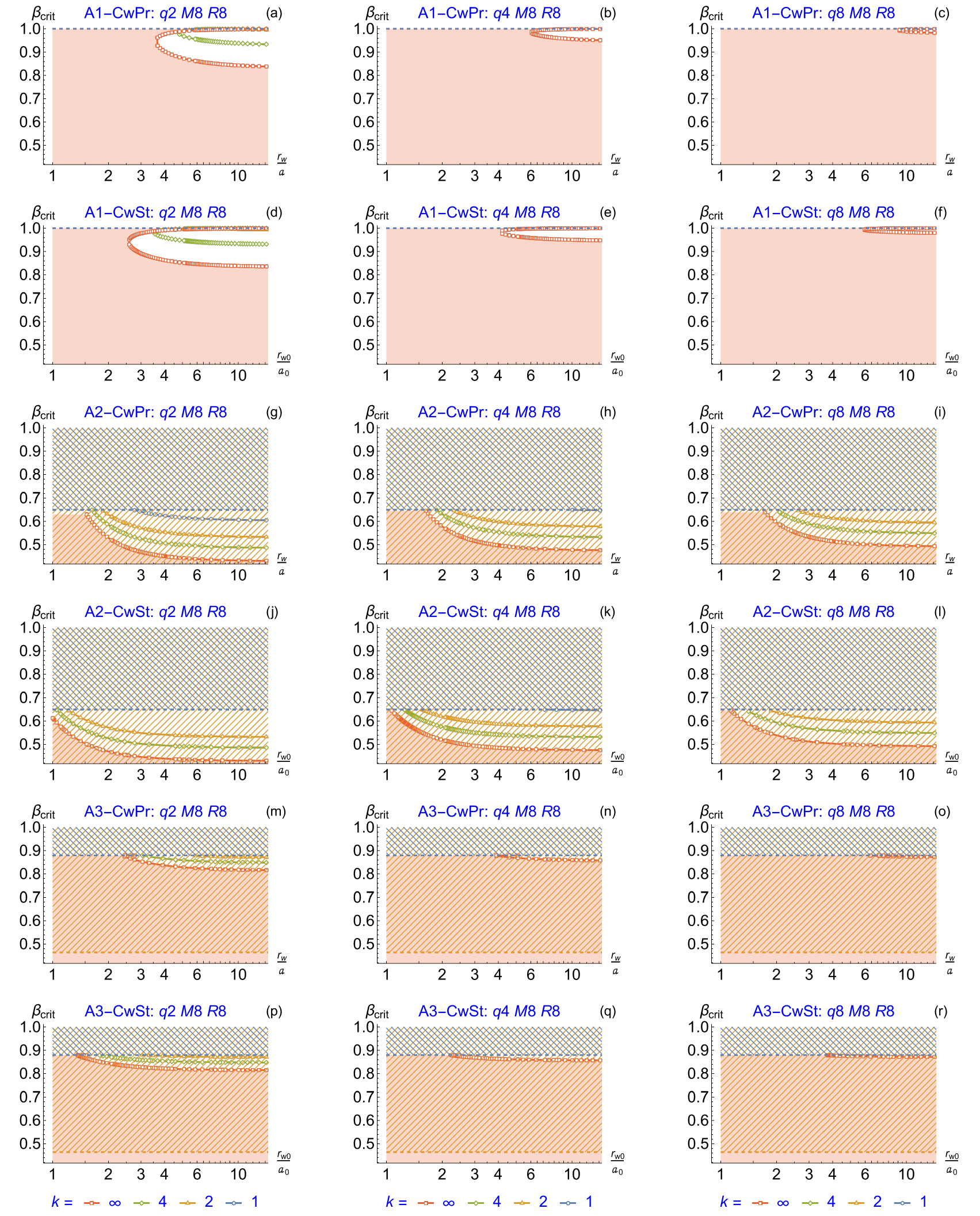}
\caption{
    Same as in Fig.~\ref{fig:M16R16} but for $M=R=8$.
}
 \label{fig:M8R8}
\end{figure*}
\begin{figure*}
\includegraphics[width=\linewidth]{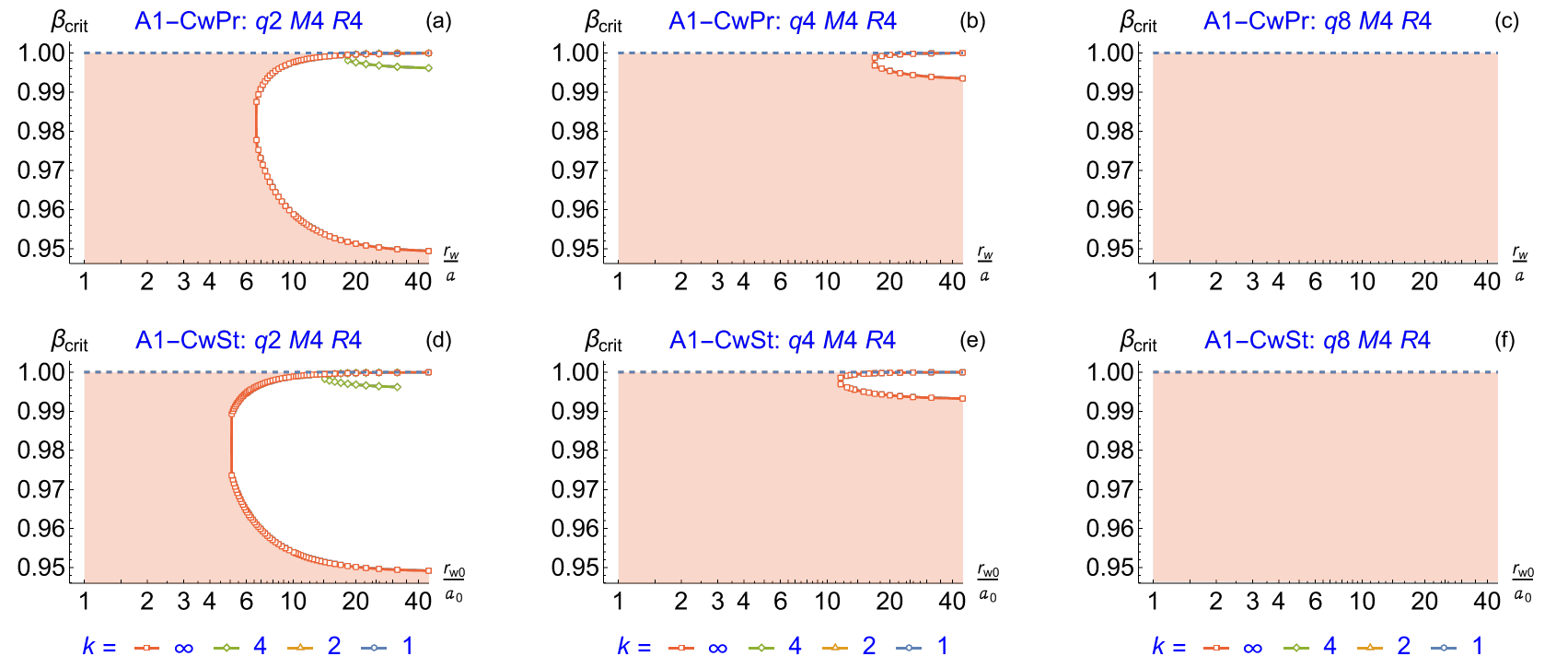}
\caption{
    Same as in Fig.~\ref{fig:M16R16} but for $M=R=4$.
    Rows for the A2 and A3 model are dropped since the rigid ballooning mode is stable in the entire region below the mirror instability threshold.
}
\label{fig:M4R4}
\end{figure*}

A series of Figures~\ref{fig:M16R16}, \ref{fig:M8R8} and ~\ref{fig:M4R4} visualizes the calculation results for the special case $R=M$ when the hot plasma component occupies the entire mirror trap from plug to plug. Initially, this case was highlighted among many others in Ref.~\citep{Kotelnikov+2020NF_60_067001} in order to compare the stability of anisotropic plasma, which is formed under transverse NBI, with the stability of isotropic plasma. In the A1 model, the $R=M$ variant corresponds to the minimum anisotropy. However, in models A2 and A3, parameter $R$ cannot serve as a measure of anisotropy, as shown in section \ref{s3.ff}, but it is related to the injection angle $\theta_{\text{inj}}$ by Eq.~\eqref{003.4:15}. Too large values of $R$,  $R\gg 2(n+1)/5$,correspond to too small angles of injection  $\theta_{\text{inj}}$, which can hardly be of interest from a practical point of view. Taking into account this explanation, the range of parameters $R$ is further reduced compared to the range that was adopted in previously published analogues for the transverse NBI model, in particular in Fig.~9\textsuperscript{\citep{Kotelnikov+2023NF_63_066027}}.

In the first two rows, Figs.~\ref{fig:M16R16}, \ref{fig:M8R8} and ~\ref{fig:M4R4} present the stability maps for the A1 model, corresponding to the transverse NBI. In Figs.~\ref{fig:M16R16} and \ref{fig:M8R8} composed respectively for $M=16$ and $M=8$, they are followed by two rows of graphs for the oblique NBI model A2 and last two rows contain maps for another oblique NBI model A3. Fig.~ \ref{fig:M4R4} has only two tows since plasma produced by oblique NBI is stable against ballooning perturbation in the entire range of $\beta$ below mirror instability threshold.

The difference between the first two rows and the four following ones is so obvious that it almost could not be commented on. With oblique injection, there is practically no upper stability zone $\beta _{\text{cr}2}$, which was present both during transverse NB injection and in isotropic plasma, as shown in figure
    10\textsuperscript{\citep{Kotelnikov+2023NF_63_066027}}
in Ref.~\citep{Kotelnikov+2023NF_63_066027}.
It can be said that with oblique NBI, the upper stability zone is absorbed in the region of mirror mode instability. 

One can also notice a significant decrease in the value of $\beta_{\text{cr}1}$ for all radial pressure profiles. This effect is accompanied by appearance of an instability zone for smooth radial profiles (with indices $k=1$, $k=2$) in the range of small values of parameter $r_{w}/a$, at which such zones were not present in the A1 model. The brown hatching from the third to sixth rows means that in the A2 and A3 models, the margin of the stability zone of the rigid ballooning mode passes entirely inside the zone of fire hose MHD instability. It is important to emphasize that the threshold of fire hose instability is determined by the criterion \eqref{003:06}, which is obtained for a homogeneous magnetic field. In an inhomogeneous magnetic field, the threshold should be slightly lower, as shown in V.~Mirnov's dissertation \citep{Mirnov1986diss(en)}. 
We are not aware of experiments with plasma in open traps in which manifestations of hose instability would be reliably recorded. Therefore, it would be premature to assert that a transition beyond the formal boundary of fire hose instability is not possible.


As for the threshold of mirror instability, the PEK code is currently unable to perform calculations in the area above its threshold.


As can be seen from the analysis of Fig.~\ref{fig:M16R16}, \ref{fig:M8R8} and~\ref{fig:M4R4}, with joint stabilization by the lateral wall and end MHD stabilizers, the instability zone $\beta _{\text{cr}1} <\beta< \beta _{\text{cr}2}$ decreases as $q$ increases. In other words, open traps with short magnetic plugs are more stable than traps in which the magnetic field increases smoothly as you move from the center of the trap to the magnetic plugs. This observation is consistent with the long-standing calculations of V. Mirnov and O. Bushkova. They calculated the threshold of ballooning instability with respect to small-scale disturbances \citep{BushkovaMirnov1986VANT_2_19e}, ignoring the effects of the finite Larmor radius. Later, the results of those calculations were confirmed in the work \citep{Kotelnikov+2021PST_24_015102}.


With fixed parameters $q$, $M$, $R$, the instability zone is maximal for the steepest radial pressure profile ($k=\infty$) and may be absent altogether for smooth profiles ($k=1$, $k=2$). Without stabilization by the end MHD anchors, the opposite situation occurs: the stability zone is smaller and may be absent altogether for smooth profiles.

\subsection{Effect of pressure peak width, $1<R<M$}
\label{s6.2}

The case $R=M$, considered in section \ref{s6.1}, corresponds to the widest pressure profile along the trap axis. Recall that parameter $R$ in the magnetic field model \eqref{03:11} makes sense of a plug ratio in the section of an open trap where the pressure of hot ions decreases to almost zero. Thus, parameter $R$ is a characteristic of the extent of the plasma pressure peak localization region, which is universally suitable for all A1, A2, A3\ldots anisotropic pressure models studied so far. In addition, in the A1 model, which simulates transverse NBI, parameter $R$ characterizes the degree of plasma anisotropy, whereas in the case of oblique NBI, described by A2 and A3 models, it is associated with the angle of inclination of the NB injection to the direction of the magnetic field.  

To study the effect of the extent of the region occupied by fast ions, which are formed during the injection of beams of neutral atoms, in real experiments at the GDT facility at the Budker Institute of Nuclear Physics, the magnetic field is reprofiled in axial direction, forming a short well \citep{Shmigelsky+2024JPP_90_975900206}. A similar operation in the numerical study with the model field \eqref{03:11} in use involves increasing parameters $q$ and $M$. As can be seen from Fig.~\ref{fig:Bv_vs_z_q}, short magnetic plugs as in the GDT correspond to higher values of $q$ and $M$ than those adopted in the current and earlier works. Corresponding calculations for model magnetic field \eqref{03:11} are planned to be published together with calculations for the real magnetic field in GDT and ALIANCE facilities.

\begin{figure*}
  \centering
\includegraphics[width=\linewidth]{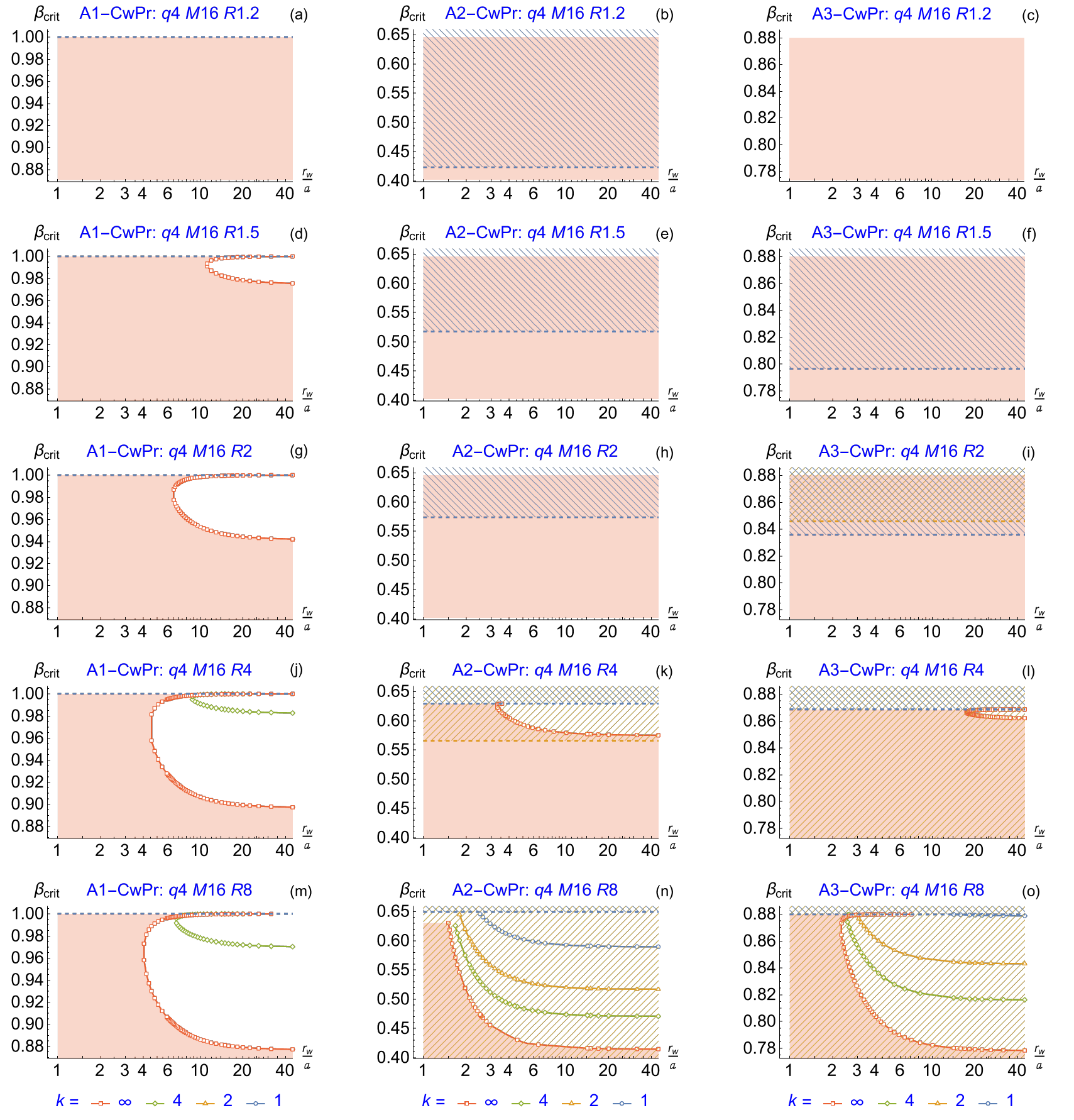}
\caption{
    Stability maps versus ratio $r_{w}/a$ for the A1, A2, and A3 pressure models in CwPr configuration simulating combined stabilization by the proportional lateral conducting chamber and end MHD anchors;
    $q=4$, $M=16$, $R\in\{1.2, 1.5, 2, 4, 8\}$.
    The RBM instability zone is located between $\beta_{\text{cr}1}(r_{w}/a)$ (the lower branch of the marginal curve) and $\beta_{\text{cr}2}(r_{w}/a)$ (the upper branch of the same color); in the case of oblique NBI, which corresponds to models A2 and A3, the upper branch is completely or partially absorbed by the mirror instability area (where the PEK fails to find it); the RBM stability zone is shaded for plasma with a sharp boundary ($k=\infty $), for which it has minimal dimensions; for large $R$, the RBM stability zone is partially located inside the hatched regions, where either the mirror, or the fire hose, or both the modes are unstable.
  }
  \label{fig:GM2-A3-CWPr_beta_vs_rw_setM16q4}
\end{figure*}

\begin{figure*}
  \centering
\includegraphics[width=\linewidth]{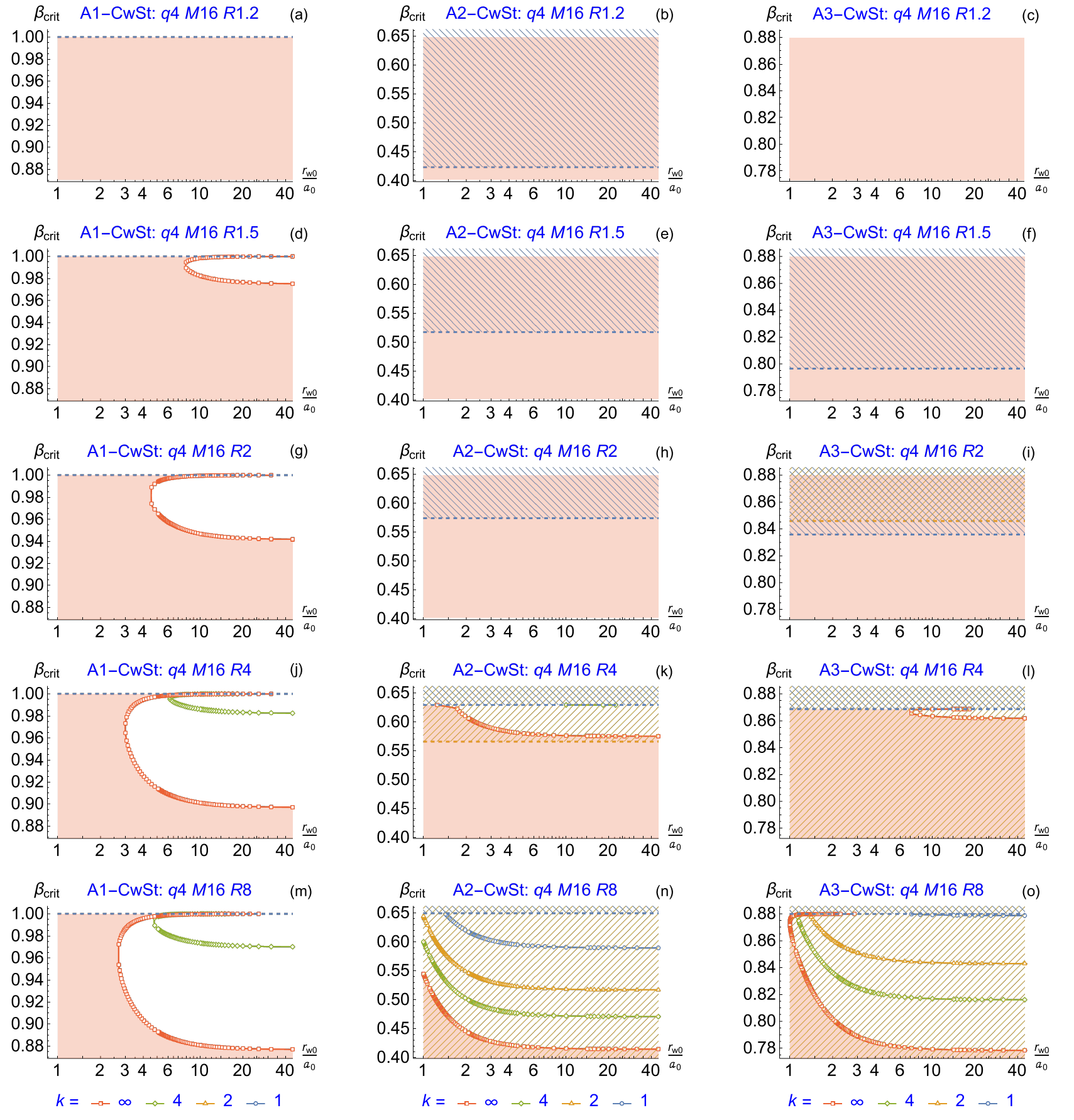}
\caption{
    Same as in Fig.~\ref{fig:GM2-A3-CWPr_beta_vs_rw_setM16q4}, but for  CwSt configuration simulating combined stabilization by the straightened lateral conducting wall and the end MHD anchors.
  }
  \label{fig:GM2-A3-CWSt_beta_vs_rw_setM16q4}
\end{figure*}

Maps of stability for the case $q=4$, $M=16$ and five values of parameter $R\in{1.2, 1.5, 2, 4, 8}$ are presented in Fig.~\ref{fig:GM2-A3-CWPr_beta_vs_rw_setM16q4} for the proportional lateral chambers and fig.~\ref{fig:GM2-A3-CWSt_beta_vs_rw_setM16q4} for the straightened  chambers. The first thing that catches eye when analyzing these figures is the absence of an unstable zone at small values of  parameter $R$ for all three models of anisotropic pressure A1, A2, A3, and the marginal value of $R$ for transverse NBI is noticeably smaller than for oblique NBI. From this point of view, oblique NBI appears to be preferable to transverse NBI.

The second conclusion that arises when comparing the graphs in same line in these figures is that transverse NBI, with other things being equal, allows one to achieve higher values of relative pressure $\beta$ than oblique NBI. This circumstance is due to the fact that the limiting beta in the case of oblique injection is limited by mirror instability, not ballooning one. From this point of view, transverse NBI appears to be preferable to oblique NBI.

\section{Conclusions}\label{s9}

This article completes a series of three publications \citep{Kotelnikov+2022NF_62_096025, Kotelnikov+2023NF_63_066027, ZengKotelnikov2024PPCF_66_075020}, in which the stability of a rigid ballooning mode with an azimuthal number $m=1$ was studied using a model magnetic field profile \eqref{03:11} along axis of a linear open trap and model pressure profiles of the hot plasma component along the radius \eqref{003.1:02}. The use of model profiles at the stage of development, debugging and testing of the PEK software package was a natural and reasonable solution, if only because as the model became more complex, the calculation time for one point on the stability map increased from tens of milliseconds in the A0 isotropic plasma model to 40 minutes in the A3 model. The use of model functions, depending on a small set of parameters, to approximate the vacuum magnetic field and the radial distribution of plasma pressure, made it possible to obtain relatively simple recommendations for achieving maximum beta in open traps. The main results can be formulated in several statements:
\begin{itemize}

  \item
    When only a lateral conducting wall is used for MHD stabilization, stable plasma confinement occurs in the upper zone $\beta > \beta _{\text{cr}2}$, which expands with a decrease in the mirror ratio $R$ at the turning point of fast ions produced by NBI. A steep radial pressure profile is the most stable in this case. The dependence on the magnetic field profile is minimal for transverse NBI, but significant for oblique NBI. Second marginal beta $\beta _{\text{cr}2}$ is only slightly affected by the mirror ratio $M$ and axial profile of the vacuum magnetic field.

  \item
    When only end MHD anchors are used, stable confinement occurs in the lower  zone $\beta < \beta _{\text{cr}1}$, which expands with a decrease in $M$ and/or $R$. Smooth radial pressure profile and steep axial magnetic field profile are the most stable. First marginal beta $\beta _{\text{cr}1}$ is strong;y affected by the mirror ratio $M$ and axial profile of the vacuum magnetic field.

  \item
    In case of oblique NBI at a small angle of injection ($R\gtrsim4$), the beta is limited by the threshold of fire hose instability rather then by the ballooning instability.

  \item
    Wall stabilization of ballooning perturbations is especially effective in combination with the end MHD anchors, which make it possible to suppress all MHD oscillations in the range $0<\beta<1$ for transverse NBI and $0<\beta<\beta_{\text{mm}}(R)$ for oblique NBI.

  \item
    The size of the stable zones in all cases essentially depends on the radial and axial profiles of plasma pressure, width and shape of the vacuum gap, plasma anisotropy, angle of injection, and stability margin of MHD anchors. The axial profile of the magnetic field is significant in all cases when the end MHD stabilizers are switched on.
\end{itemize}

Further development of the PEK software package should include a series of minor and major upgrades. The first step in this queue will be to connect the module for calculating the actual magnetic field in the GDT, CAT and ALIANCE facilities. At the program level, this step has already been completed. The other most anticipated major modernization should be the inclusion of dissipative processes, both inside the plasma and in the conductors surrounding it.

\section*{Acknowledgments}
    
    The author thanks Peter Bagryansky (BINP), 
    Cary Forest (UW-Madison),
    Alexander Ivanov (TAE Technologies), 
    Vadim Prikhodko (BINP), Sergey Putvinsky (TAE Technologies), Evgeny Shmigelsky (BINP), Dmitry Yakovlev (UW-Madison), Peter Yushmanov (TAE Technologies), and Qiusun Zeng (HFIPS) for numerous discussions and valuable comments at different stages of development of the PEK package.
    
    Special thanks to Alexey Beklemishev (BINP) for discussing the conditions for the applicability of the LoDestro equation, Victor Ilgisonis (ROSATOM) for discussing the fire hose instability threshold, Alexander Milstein (BINP) for discussing the properties of second-order ordinary differential equations with singular coefficients, and Vladimir Mirnov (UW-Madison) for answering many of my questions.

\section*{Funding} 

    This work was supported by the Russian Science Foundation under the Grant 24-12-00309 awarded to Novosibirsk State University [Sections \ref{s05} and \ref{s06}].

    This work is also a part of the state assignment of the Russian Federation for the Budker Institute of Nuclear Physics  [Sections \ref{s02}–\ref{s04}].

\section*{Supplementary data}

No supplementary material and movies are available.

\section*{Data availability statement}

    The data that support the findings of this study are not openly available..

\section*{Declaration of interests}


The author reports no conflict of interest.

\section*{Author ORCID}

\noindent
Igor KOTELNIKOV \href{https://orcid.org/0000-0002-5509-3174}{https://orcid.org/0000-0002-5509-3174}

\appendix

\section{LoDestro equation}\label{A1}

The LoDestro equation is a second-order ordinary differential equation for the function
    \begin{equation}
    \label{A2:02}
    \phi(z) = a(z) B_{v}(z) \xi_{n}(z)
    ,
    \end{equation}
which depends on single coordinate $z$ along the trap axis and is expressed in terms of the variable radius of the plasma column boundary $a=a(z)$, the vacuum magnetic field $B_{v}=B_{v}(z)$ and the virtual small displacement $ \xi_{n}=\xi_{n}(z)$ of the plasma column from the axis.
In its final form, the LoDestro equation reads
    \begin{multline}
    \label{A2:01}
    0 = \der{}{z}
    \left[
        \Lambda + 1 - \frac{2\mean{\overline{p}}}{B_{v}^{2}}
    \right]
    \der{\phi}{z}
    \\
    +
    \phi
    \left[
        - \der{}{z}\left(
            \frac{B_{v}'}{B_{v}} + \frac{2a'}{a}
        \right)
        \left(
            1 - \frac{\mean{\overline{p}}}{B_{v}^{2}}
        \right)
    +
    \frac{\omega^{2}\mean{\rho}}{B_{v}^{2}}
    \right.
    \\
    \left.
    -
    \frac{2\mean{\overline{p}}}{B_{v}^{2}}\frac{a_{v}''}{a_{v}}
    -
    \frac{1}{2}\left(
            \frac{B_{v}'}{B_{v}} + \frac{2a'}{a}
    \right)^{2}
    \left(
        1 - \frac{\mean{\overline{p}}}{B_{v}^{2}}
    \right)
    \right]
    ,
    \end{multline}
where the derivative $\tder{}{z}$ in the first two lines acts on all factors to the right of it, and the prime ($'$) is a shortcut for $\tder{}{z}$.
%
Other notations are defined as follows
    \begin{gather}
    \label{A2:03}
    \frac{a^{2}}{2} = \int_{0}^{1} \frac{\dif{\psi}}{B}
    ,\\
    \label{A2:03a}
    \frac{r^{2}}{2} = \int_{0}^{\psi} \frac{\dif{\psi}}{B}
    ,\\
    \label{A2:04}
        B^{2} = B_{v}^{2} -2p_{\bot}
    ,\\
    \label{A2:07}
    a_{v}(z) = \sqrt{\frac{2}{B_{v}(z)}}
    ,\\
    \label{A2:05}
    \overline{p} = \frac{p_{\bot} + p_{\|}}{2}
    ,\\
    \label{A2:06}
    \mean{\overline{p}}
    =
    \frac{2}{a^{2}}
    \int_{0}^{1} \frac{\dif{\psi}}{B}\,\overline{p}
    ,\\
    \label{A2:09}
    \Lambda = \frac
    {
        r_{w}^{2} + a^{2}
    }{
        r_{w}^{2} - a^{2}
    }
    .
    \end{gather}
Equation \eqref{A2:03a} relates the radial coordinate $r$ and the magnetic flux $\psi$ through a ring of radius $r$ in the $z$ plane. The magnetic field $B=B(\psi,z)$, weakened by the plasma diamagnetism, in the paraxial (long-thin) approximation (which assumes small curvature of field lines) is related to the vacuum magnetic field $B_{v}=B_ {v }(z)$ by the equation of transverse equilibrium equation \eqref{A2:04}. Kinetic theory predicts (see, for example, \citep{Newcomb1981JPP_26_529}) that the transverse and longitudinal plasma pressures can be considered as functions of the magnetic field $B$ and magnetic flux $\psi$, i.e.\ $p_{\bot}=p_{\bot}(B ,\psi)$, $p_{\|}=p_{\|}(B,\psi)$. In Eq.~\eqref{A2:01}, one must assume that the magnetic field $B$ is already expressed in terms of $\psi$ and $z$, and therefore $p_{\bot}=p_{\bot}(\psi,z)$, $p_{\|}=p_{\|}(\psi,z)$. The angle brackets in Eq.~\eqref{A2:01} denote the mean value of an arbitrary function of $\psi$ and $z$ over the plasma cross section. In particular, the average value $\mean{\rho}$ of the density $\rho=\rho(\psi,z)$ is calculated using a formula similar to \eqref{A2:06}, and $\omega $ is the angular frequency of oscillations.

Parameter $\Lambda $ is, generally speaking, a function of $z$ coordinate. It implicitly depends on the plasma parameters and  magnetic field through the dependence of the plasma column radius $a=a(z)$ on them. In the special case of proportional chamber, when $r_{w}(z)/a(z)=\const$, the function $\Lambda(z)$ becomes  constant, which simplifies the equation somewhat. Namely, for the sake of such simplification, in most papers of other authors they assume that $\Lambda =\const$.

Traditionally, two types of boundary conditions are considered. In the presence of conducting end plates located in the magnetic mirrors at $z=\pm1$, the boundary condition
     \begin{equation*}
     \phi(\pm1) = 0
     \end{equation*}
should be chosen. In a more general case, when the conducting end plate is installed somewhere in the behind-the-mirror region, namely, in the plane with coordinates $z=\pm z_{E}$, the zero boundary condition must obviously be assigned to this plane:
     \begin{equation}
     \label{A2:11S}
     \phi(\pm z_{E}) = 0
     .
     \end{equation}
By solving the LoDestro equation with the boundary condition \eqref{A2:11S}, it is possible to simulate the effect of the end MHD anchors with different stability margins.

If the plasma ends are electrically isolated, the boundary condition
    \begin{equation}
    \label{A2:12}
    \phi'(\pm z_{E}) = 0
    \end{equation}
is applied at $z=\pm z_{E}$. As a rule, it implies that other methods of MHD stabilization in addition to stabilization by a conducting lateral wall are not used.


An obvious fit to the LoDestro equation with boundary conditions \eqref{A2:11S} or \eqref{A2:12} is the trivial solution $\phi\equiv0$. To eliminate the trivial solution, we impose a normalization in the form of one more condition
    \begin{equation}
    \label{A2:14}
    \phi(0)=1.
    \end{equation}
%
Taking into account the symmetry of the magnetic field in actually existing open traps with respect to the median plane $z=0$, it suffices to find a solution to the LoDestro equation at half the distance between the magnetic mirrors, for example, in the interval $0<z<1$. Due to the same symmetry, the desired function $\phi(z)$ must be even, therefore
    \begin{equation}
    \label{A2:15}
    \phi'(0)=0.
    \end{equation}
It is convenient to search for a solution to the LoDestro equation by choosing the boundary conditions \eqref{A2:14} and \eqref{A2:15}. In theory, a second-order linear ordinary differential condition with two boundary conditions must always have a solution. However, the third boundary condition \eqref{A2:11S} or \eqref{A2:12} can only be satisfied for a certain combination of parameters. If the parameters of the plasma, magnetic field, and geometry of the lateral conducting wall are given, the third boundary condition should be considered as a nonlinear equation for the squared frequency $\omega^{2}$ of frequency $\omega$. If the root of such an equation is positive, the MHD oscillations with azimuthal number $m=1$ are stable; if $\omega ^{2}<0$, then instability takes place. On the margin of the stability zone $\omega ^{2}=0$. In this case, the solution of the boundary-value problem \eqref{A2:01}, \eqref{A2:14}, \eqref{A2:15} with the additional boundary condition \eqref{A2:11S} or \eqref{A2:12} gives the critical value of beta, respectively both $\beta_{\text{cr}1}$ and $\beta_{\text{cr}2}$ or only $\beta_{\text{cr}2}$.

As it was mentioned by \citet{Kotelnikov+2023NF_63_066027}, the LoDestro equation \eqref{A2:01} with boundary conditions \eqref{A2:11S} or \eqref{A2:12} constitutes the standard Sturm-Liouville problem. At first glance, it may seem that the solution of such a problem is rather standard. However, the equation has the peculiarity that its coefficients could be singular. In the anisotropic pressure model A1 relevant to the transverse NBI, the singularity appears near the minimum of the magnetic field in the limit $\beta \to 1$. In case of oblique NBI models A2, A3, \ldots , some coefficients of the LoDestro equations can have singularity near point $z$ where the condition \eqref{003:07} of the mirror stability breaks at the plasma column axis. 

For example, in the A3 model, the singularity occurs at $p_{0} \to p_{\text{mm}}=4$ in a magnetic field $b=3/4$, which corresponds to $b_{v}=\sqrt{27/32}$. Specifically, the coefficient at the function $\phi(z)$ in equation \eqref{A2:01} has a second-order pole, $-{217 b_{v}'(z)^2}/{36 (p-p_{\text{mm}})^{2}}$, and the coefficient at the first derivative $\phi'(z)$ has a first-order pole, $11b_{v}'(z)/9\sqrt{6}(p-p_{\text{mm}})$. For the magnetic field model \eqref{03:11}, the coordinate of the singular point can be found explicitly:
    \begin{gather}
    \label{A2:15a}
    z_{\text{mm}}
    =
    \frac{2}{\pi }
    \sin^{-1}\left(
        \left(
            \frac{3 \sqrt{6} R-8}{8M-8}
        \right)^{{1}/{q}}
    \right)
    .
    \end{gather}

Reproducing the method of Section~5 of Ref.~\citep{Kotelnikov+2023NF_63_066027}, we denote by $z_{R}$ the coordinate of the turning point on the $z$ axis, where $b=b_{v}=1 $ and $p_{\bot} =p_{\|}=0$. For the magnetic field model \eqref{03:11} we have
    \begin{equation}
    \label{A2:16}
    z_{R}
    =
    \frac{2}{\pi }\,
    \arcsin\left(
        \left(
            \frac{R-1}{M-1}
        \right)^{{1}/{q}}
    \right)
    .
    \end{equation}
In the adopted models of anisotropic plasma, its pressure is zero in the region $z_{R}<z<1$ between the turning point $b=b_{v}=1$ and the magnetic mirror throat $b=b_{v}=M/R$.  At zero pressure, the LoDestro equation \eqref{A2:01} takes an extremely simple form
    \begin{equation}
    \label{A2:17}
    0 = \der{}{z}
    \left[
        \left(
            \Lambda + 1
        \right)
        \der{\phi}{z}
    \right]
    .
    \end{equation}
Its solution is the equality
    \begin{equation}
     \label{A2:18}
    \left[
        \Lambda(z) + 1
    \right]
    \phi'(z)
    =\const,
    \end{equation}
where the constant on its right-hand side can be found from the boundary condition at $z=z_{E}$.

In case of plasma with electrically isolated ends, the boundary condition \eqref{A2:12} at $z=z_{E}$ should be applied, so that $\phi'(z_{E})=0$. Hence, the constant on the right-hand side of Eq.~\eqref{A2:18} is also zero. Since the factor $\Lambda(z) + 1$ is greater than zero everywhere, we conclude that $\phi'(z)=0$ in the entire region $z_{R} < z < z_{E}$. Thus, it is sufficient to find a numerical solution of the original equation \eqref{A2:01} in the region $0 < z < z_{R}$, but one must be careful when setting the boundary conditions for $z=z_{R} $.

It should be taken into account that the derivative $\phi'(z)$ undergoes a jump at $z=z_{R}$. Indeed, integrating Eq.~\eqref{A2:01} over an infinitesimal neighborhood of the point $z_{R}$ from $z_{R}^{-}$ to $z_{R}^{+}$ yields the equation
    \begin{equation}
    \label{A2:19}
    \left[ \Lambda(z_{R}) + 1 \right]
    \left[
        \phi'({z_{R}^{+}})
        -
        \phi'({z_{R}^{-}})
    \right]
    =
    \left[
        Q({z_{R}^{+}})
        -
        Q({z_{R}^{-}})
    \right]
    \phi(z_{R})
    ,
    \end{equation}
in which we took into account that $\Lambda $, $\phi$ and $\mean{\overline{p}}$ are continuous at the point $z=z_{R}$, in contrast to the derivative $\phi'(z)$ and the coefficient
    \begin{equation}
    \label{A2:20}
    Q(z) =
    \frac{B_{v}'}{B_{v}}
    +
    \frac{2a'}{a}
    =
    \frac{2a'}{a}
    -
    \frac{2a_{v}'}{a_{v}}
    .
    \end{equation}
The jump in the coefficient is due to the fact that for $b=b_{v}=1$ the derivative of functions such as \eqref{003.2:01} jumps (only in case of pressure models A1 and A2). Since $\phi'({z_{R}^{+}})=0$ and $Q({z_{R}^{+}})=0$, Eq.~\eqref{A2:19} yields the value $\phi'(z_{R}^{-})$ that the derivative of $\phi'(z)$ must have at the point $z_{R}^{-}$ on the right boundary of the interval $0<z<z_{R}$ from its inner side:
    \begin{equation}
    \label{A2:21}
    \phi'({z_{R}^{-}})
    =
    \frac{
        Q({z_{R}^{-}})
    }{
        \Lambda(z_{R}) + 1
    }\,
    \phi(z_{R})
    .
    \end{equation}
When solving Eq.~\eqref{A2:01} on the interval $0<z<z_{R}^{-}$, the boundary condition \eqref{A2:21} should be used instead of \eqref{A2:12}. Note that the $z_{E}$ coordinate is not included in \eqref{A2:21}.

For all pressure models parameter $Q({z_{R}^{-}})$ can be found in analytic form.
Namely, for model A1,
    \begin{gather*}
    Q_{\infty}^{-} 
    = 
    -
    \frac{2 p}{(1-2 p)}
    b_{v}'(1)
    ,
    \\
    Q_{1}^{-} 
    = 
    \left(
        1 + \frac{\log (1-2 p)}{2p}
    \right)
    b_{v}'(1)
    ,\\
    Q_{2}^{-} = 
    \frac{1}{2} \left(\frac{\sqrt{2} \sqrt{\frac{1}{p}-2} \sin^{-1}\left(\sqrt{2} \sqrt{p}\right)}{2 p-1}+2\right)
    b_{v}'(1)
    ,
    \end{gather*}
    \begin{multline*}
    Q_{4}^{-} 
    = 
    \frac{1}{2} 
    \left(
        \left(\frac{2p}{2 p-1}\right)^{3/4} 
        \left(
            \tan^{-1}
            \left(
                \sqrt[4]{\frac{2p}{2 p-1}}
            \right)
    \right.
        \right.
        \\
        \left.
    \left.
            +
            \tanh^{-1}
            \left(
                \sqrt[4]{\frac{2p}{2p-1}}
            \right)
        \right)
    \right.
    \\
    \left.
        +
        \left\{
            8\sqrt[4]{p}+\sqrt[4]{2-4 p} 
            \log\left(
                \sqrt{1-2 p}
                +
                \sqrt{2p}
                -
                2^{3/4}\sqrt[4]{(1-2 p) p
                }
            \right)
    \right.
        \right.
        \\
        \left.
    \left.
            -
            \sqrt[4]{2-4 p}
            \log\left(
                \sqrt{1-2 p}
                +
                \sqrt{2p}
                +
                2^{3/4}\sqrt[4]{(1-2p)p}
            \right)
    \right.
        \right.
        \\
        \left.
    \left.
            -
            2 (-1)^{3/4} \sqrt[4]{4p-2} 
            \tan^{-1}\left(
                1-\frac{1+i}{\sqrt[4]{1-\frac{1}{2 p}}}
            \right)
    \right.
        \right.
        \\
        \left.
    \left.
            +
            2(-1)^{3/4} \sqrt[4]{4 p-2} 
            \tan^{-1}\left(
                1+\frac{1+i}{\sqrt[4]{1-\frac{1}{2 p}}}
            \right)
        \right\}
        /
        \left\{
            4\sqrt[4]{p}
        \right\}
    \right)
    b_{v}'(1)
    ;
    \end{multline*}
for model A2,
    \begin{gather*}
    Q_{\infty}^{-} 
    = 
    \frac{p}{p-1}
    b_{v}'(1)
    ,\\
    Q_{1}^{-} = \frac{p+\log (1-p)}{p}
    b_{v}'(1)
    ,\\
    Q_{2}^{-} 
    = 
    \frac{1}{2} 
    \left(
        2-\frac{2 \sin^{-1}\left(\sqrt{p}\right)}{\sqrt{(1-p) p}}
   \right)
    b_{v}'(1)
    ,
    \\
    Q_{4}^{-} 
    = 
    \frac{1}{2} 
    \left\{
        \left[
            8 \sqrt[4]{p}
    \right.
        \right.
        \\
        \left.
    \left.
            +
            \sqrt{2} \sqrt[4]{1-p} \log
            \left(
                \sqrt{1-p}+\sqrt{p}-\sqrt{2} \sqrt[4]{(1-p) p}
            \right)
    \right.
        \right.
        \\
        \left.
    \left.
            -
            \sqrt{2} \sqrt[4]{1-p} 
            \log\left(
                \sqrt{1-p}+\sqrt{p}+\sqrt{2} \sqrt[4]{(1-p) p}
            \right)
    \right.
        \right.
        \\
        \left.
    \left.
            +
            2 \sqrt{2} \sqrt[4]{1-p} 
            \tan^{-1}\left(
                1-\sqrt{2}\sqrt[4]{\frac{p}{1-p}}
            \right)
    \right.
        \right.
        \\
        \left.
    \left.
            -
            2 \sqrt{2} \sqrt[4]{1-p} 
            \tan^{-1}\left(
                \sqrt{2} \sqrt[4]{-\frac{p}{p-1}}+1
            \right)
        \right]
        /
        {
        \left[
            4\sqrt[4]{p}
        \right]
        }
        \right.
        \\
        \left.
        +
        \left(
            \frac{p}{p-1}\right)^{3/4} 
            \left(
                \tan^{-1}\left(\sqrt[4]{\frac{p}{p-1}}\right)
                +
                \tanh^{-1}\left(\sqrt[4]{\frac{p}{p-1}}\right)
            \right)
    \right\}
    b_{v}'(1)
    ,
    \end{gather*}
where
    \begin{gather*}
    b_{v}'(1)
    =
    \frac{\pi  q (R-1) }{2R}\,
    \sqrt{\left(\frac{M-1}{R-1}\right)^{2/q}-1}
    .
    \end{gather*}
For model A3 and above,
    \begin{gather*}
    Q_{\infty}^{-} = Q_{1}^{-} = Q_{2}^{-} = Q_{4}^{-} = 0
    .
    \end{gather*}

Let's move on to solving Eq.~\eqref{A2:01} with boundary conditions \eqref{A2:14} and \eqref{A2:15} for $z=0$ and \eqref{A2:11S} for $z=z_{E}$ and. As shown above, in the region $z_{R}<z<z_{E}$ the desired solution satisfies Eq.~\eqref{A2:18} but the constant in this equation is now equal to
    $
    \left[
        \Lambda(z_{R}) + 1
    \right]
    \phi'(z_{R}^{+})
    $,
rather than zero. Hence, the derivative $\phi'(z)$ at $z=z_{R}^{+}$ is equal to
    \begin{gather}
    \label{A2:31}
    \phi'(z_{R}^{+})
    =
    \frac{- \phi(z_{R})}
    {
        \left[ \Lambda(z_{R}) + 1 \right]
        \int_{z_{R}}^{z_{E}}\frac{\dif{z}}{\Lambda(z) + 1}
    }
    .
    \end{gather}
Substituting $\phi(z_{R}^{+})$ into Eq.~\eqref{A2:19} and taking into account that $Q({z_{R}^{+}})=0$, yields the derivative $\phi'({z_{R}^{-}})$ on the right boundary of the interval $0<z<z_{R}$ from its inner side:
    \begin{gather}
    \label{A2:32}
    \phi'({z_{R}^{-}})
    =
    \left[
        \frac{Q({z_{R}^{-}})}{
            \Lambda(z_{R}) + 1
            \vphantom{\int_{z_{R}}^{z_{E}}}
        }
        -
        \frac{1}
        {
            \left[ \Lambda(z_{R}) + 1 \right]
            \int_{z_{R}}^{z_{E}}\frac{\dif{z}}{\Lambda(z) + 1}
        }
    \right]
    \phi(z_{R})
    .
    \end{gather}
This is the boundary condition that should be used instead of \eqref{A2:11S} in the problem of ballooning instability with a combination of lateral wall stabilization and stabilization by conducting end plates. In the case of $\Lambda(z)=\const$ under consideration in this paper it reduces to
    \begin{equation}
    \label{A2:33}
    \phi'({z_{R}^{-}})
    =
    \left[
    \frac{
        Q({z_{R}^{-}})
    }{
        \Lambda + 1
    }
    -
    \frac{
        1
    }{
        z_{E} - z_{R}
    }
    \right]
    \phi(z_{R})
    .
    \end{equation}

If a conducting limiter is installed before the magnetic plug at mirror ratio $L$, $R<L<M$, as in  the RwPr or RwSt configurations, then the limiter coordinate $z_{E}$ is calculated by a formula similar to \eqref{A2:16}, in which $R\to L$ should be replaced. If the conducting end plate is placed behind the magnetic plug as in BwSt or BwPr configuration, then
    \begin{gather}
    z_{E} = 2
    -
    \frac{2}{\pi }\,
    \arcsin\left(
        \left(
            \frac{L-1}{M-1}
        \right)^{{1}/{q}}
    \right)
    .
    \end{gather}
The integrals in Eqs.~\eqref{A2:31} and \eqref{A2:32} were calculated analytically. For example, for the magnetic field model \eqref{003.3:01} with the index $q=2$, a relatively short expression is found
    \begin{multline*}
    \int_{z_{R}}^{1}
    \frac{\Lambda(z_{R}) + 1}{\Lambda(z) + 1}
    \dif{z}
    =
    \frac{
        1
    }{
        \pi\,\sqrt{M} \left(r_w^2-2\right)
    }
    \\
    \times
    \left[
        -\pi\left(\sqrt{M}\, r_w^2 \left(z_s-1\right)+M+1\right)
        +
        \pi\,(M-1) \cos\left(\pi z_s\right)
    \right.
    \\
    \left.
        +
        2 \left(
            M+1
            -
            (M-1) \cos\left(\pi z_s\right)
        \right)
        \tan ^{-1}\left(
            \sqrt{M}
            \tan \left(\frac{\pi  z_s}{2}\right)
        \right)
    \right]
    ,
    \end{multline*}
where
    \begin{gather*}
    r_{w}^{2}
    =
    a_{0}^{2}\,
    \frac{\Lambda_{0}+1}{\Lambda_{0}-1}
    >
    2
    \end{gather*}
denotes the dimensionless square of the radius of the conducting wall in the model of straightened cylinder. For magnetic field profiles with indices $q=4$ and $q=8$, much more cumbersome formulas are obtained, but they still speed up the calculation of integrals by several times as compared to numerical integration.

\bibliographystyle{jpp}

\bibliography{GDMT}

\end{document}